\documentclass[11pt]{article}
\usepackage{a4wide}
\input epsf

\newcommand{\epostfig}[3]{
\begin{figure}[tbp]
\setlength{\epsfxsize}{1.25\hsize}
\hspace*{-0.1\hsize} \epsfbox{#1}
\caption{\label{#2}#3}
\end{figure}
}

\newcommand{\lvlv}{\mbox{$\rm \ell\nu\ell\nu$}}
\newcommand{\qqlv}{\mbox{$\rm q\bar{q}\ell\nu$}}
\newcommand{\qqev}{\mbox{$\rm q\bar{q}e\nu$}}
\newcommand{\qqmv}{\mbox{$\rm q\bar{q}\mu\nu$}}
\newcommand{\qqtv}{\mbox{$\rm q\bar{q}\tau\nu$}}
\newcommand{\qqqq}{\mbox{$\rm q\bar{q}q\bar{q}$}}
\newcommand{\bqqlv}{\boldmath{${\rm\bf q\bar{q}}\ell\nu$}}
\newcommand{\bqqqq}{$\rm\bf q\bar{q}q\bar{q}$}
\newcommand{\zb}{\mbox{$\rm Z$}}
\newcommand{\ww}{\mbox{$\rm W^+W^-$}}
\newcommand{\zz}{\mbox{$\rm ZZ$}}
\newcommand{\zg}{\mbox{$\zb/\gamma$}}
\newcommand{\epem}{\mbox{$\rm e^+e^-$}}
\newcommand{\mpmm}{\mbox{$\mu^+\mu^-$}}
\newcommand{\qqbar}{\mbox{$\rm q\bar{q}$}}
\newcommand{\ffbar}{\mbox{$\rm f\bar{f}$}}
\newcommand{\mean}[1]{\langle{#1}\rangle}

\newcommand{\mw}{\mbox{$m_{\rm W}$}}
\newcommand{\gw}{\mbox{$\Gamma_{\rm W}$}}
\newcommand{\mz}{\mbox{$m_{\rm Z}$}}
\newcommand{\mwz}{\mbox{$m^0_{\rm W}$}}
\newcommand{\gwz}{\mbox{$\Gamma^0_{\rm W}$}}
\newcommand{\ipb}{\mbox{$\rm pb^{-1}$}}
\newcommand{\sexp}{\mbox{$\sigma_{\rm exp}$}}
\newcommand{\valp}{\mbox{$\vec{\alpha}$}}
\newcommand{\mvc}{m_{\rm 5C}}

\newcommand{\mta}{\mbox{$m'_1$}}
\newcommand{\mtb}{\mbox{$m'_2$}}
\newcommand{\ma}{\mbox{$m_1$}}
\newcommand{\mb}{\mbox{$m_2$}}
\newcommand{\mac}{\mbox{$m^c_1$}}
\newcommand{\mbc}{\mbox{$m^c_2$}}

\newcommand{\levt}{{\cal{L}}^{\rm evt}}
\newcommand{\ltot}{{\cal{L}}^{\rm tot}}
\newcommand{\lsig}{{\cal{L}}^{\rm sig}}
\newcommand{\lzz}{{\cal{L}}^{\rm ZZ}}
\newcommand{\lzg}{{\cal{L}}^{\rm Z/\gamma}}
\newcommand{\psig}{{\cal{P}}^{\rm sig}}
\newcommand{\pzz}{{\cal{P}}^{\rm ZZ}}
\newcommand{\pzg}{{\cal{P}}^{\rm Z/\gamma}}

\newcommand{\rs}{\mbox{$\sqrt{s}$}}
\newcommand{\rsp}{\mbox{$\sqrt{s'}$}}

\newcommand{\syaiii}{\mbox{$p_{2.5}$}}

\newcommand{\syavi}{\mbox{$\kappa_{-0.5}$}}

\newcommand{\syaxiv}{\mbox{$\kappa_{-0.75}$}}
\newcommand{\dmjx}{\mbox{$\Delta m(J_X,\syavi)$}}
\newcommand{\dmyx}[1]{\mbox{$\Delta m(#1,\syavi)$}}
\newcommand{\dmjz}{\mbox{$\Delta m(J_X,J_0)$}}
\newcommand{\dmyz}[1]{\mbox{$\Delta m(#1,J_0)$}}

\newcommand{\precon}{\mbox{$p_{\rm rec}$}}
\newcommand{\PLB}[3] {Phys.~Lett.\ {B#1} (#2) #3}
\newcommand{\PRL}[3] {Phys.~Rev.\ {Lett.~#1} (#2) #3}
\newcommand{\PRD}[3] {Phys.~Rev.\ {D#1} (#2) #3}
\newcommand{\PRP}[3] {Phys.~Rep.\ {#1} (#2) #3}
\newcommand{\NIM}[3] {Nucl.~Instrum.\ {Methods~#1} (#2) #3}
\newcommand{\NPB}[3] {Nucl.~Phys.\ {B#1} (#2) #3}
\newcommand{\CPC}[3] {Comp.~Phys.\ {Comm.~#1} (#2) #3}
\newcommand{\ZPC}[3] {Z.~Phys.\ {C#1} (#2) #3}

\newcommand{\EPJ}[3] {Eur.~Phys.\ J.\ {C#1} (#2) #3}

\newcommand{\JHEP}[3] {JHEP {#1} (#2) #3}
\newcommand{\etal} {et~al.}
\newcommand{\mwqqlv}{80.449}
\newcommand{\mwqqqq}{80.353}
\newcommand{\mwqqjz}{80.394}
\newcommand{\mwnonl}{80.416}

\newcommand{\mwallo}{80.415}
\newcommand{\mwqqlvstat}{0.056}
\newcommand{\mwqqqqstat}{0.060}
\newcommand{\mwqqjzstat}{0.051}
\newcommand{\mwnonlstat}{0.042}

\newcommand{\mwallostat}{0.042}

\newcommand{\mwnonlsyst}{0.031}

\newcommand{\mwallosyst}{0.030}

\newcommand{\mwnonllep}{0.009}

\newcommand{\mwallolep}{0.009}
\newcommand{\mwqqlvsytt}{0.028}
\newcommand{\mwqqqqsytt}{0.058}
\newcommand{\mwqqjzsytt}{0.133}
\newcommand{\mwnonlsytt}{0.032}
\newcommand{\gwqqlv}{1.927}
\newcommand{\gwqqqq}{2.125}
\newcommand{\gwnonl}{1.996}
\newcommand{\gwqqlvstat}{0.135}
\newcommand{\gwqqqqstat}{0.112}
\newcommand{\gwnonlstat}{0.096}

\newcommand{\gwnonlsyst}{0.102}

\newcommand{\gwnonllep}{0.003}
\newcommand{\gwqqlvsytt}{0.091}
\newcommand{\gwqqqqsytt}{0.177}
\newcommand{\gwnonlsytt}{0.102}
\newcommand{\dmwc}{-0.097}
\newcommand{\dmwcstat}{0.082}
\newcommand{\dmwcsyst}{0.039}
\newcommand{\dgwc}{\hspace{3mm}0.198}
\newcommand{\dgwcstat}{0.175}
\newcommand{\dgwcsyst}{0.124}
\newcommand{\dmres}{-152}
\newcommand{\dmstat}{68}
\newcommand{\dmsyst}{41}
\newcommand{\dmbec}{45}
\begin{document}
\begin{titlepage}
{\center\Large

EUROPEAN ORGANIZATION FOR NUCLEAR RESEARCH \\

}
\bigskip

{\flushright
OPAL PR\,410 \\
CERN-PH-EP/2005-038\\
15\,th July 2005 \\
}
\bigskip 

\begin{center}
    \LARGE\bf\boldmath
Measurement of the mass and width of the W boson
\end{center}
\vspace{2mm}
\bigskip

\begin{center}
\Large The OPAL Collaboration \\
\bigskip
\end{center}
\vspace{3mm}

\begin{abstract}
The mass and width of the W boson are measured 
using $\epem\rightarrow\ww$ events from the data sample collected 
by the OPAL experiment at LEP at 
centre-of-mass energies between 170\,GeV and 209\,GeV.
The mass (\mw) and width (\gw) are determined using direct reconstruction 
of the kinematics of $\ww\rightarrow\qqlv$ and $\ww\rightarrow\qqqq$ events.
When combined with previous OPAL measurements using $\ww\rightarrow\lvlv$
events and the dependence on \mw\ of the WW production cross-section at 
threshold, the results are determined to be
\begin{eqnarray*}
\mw & = & \mwallo \pm \mwallostat \pm \mwallosyst \pm \mwallolep \rm\,GeV \\
\gw & = &  \gwnonl \pm \gwnonlstat \pm \gwnonlsyst \pm \gwnonllep \rm\,GeV
\end{eqnarray*}
where the first error is statistical, the second systematic and the 
third due to uncertainties in the value of the LEP beam energy. 
By measuring \mw\ with several different jet algorithms in the \qqqq\ channel,
a limit is also obtained on possible final-state interactions due to 
colour reconnection effects in $\ww\rightarrow\qqqq$ events.
The consistency of the results for the W mass and width with those 
inferred from other electroweak parameters
provides an important test of the Standard Model of electroweak interactions.
\end{abstract}

\begin{center}
\bigskip

\large
This paper is dedicated to the memory of Steve O'Neale\\

\vspace{7mm}

Submitted to Eur.\ Phys.\ J.\ C.

\vspace{5mm}


\end{center}

\end{titlepage}

\begin{center}{\Large        The OPAL Collaboration
}\end{center}\bigskip
\begin{center}{
G.\thinspace Abbiendi$^{  2}$,
C.\thinspace Ainsley$^{  5}$,
P.F.\thinspace {\AA}kesson$^{  3,  y}$,
G.\thinspace Alexander$^{ 22}$,
G.\thinspace Anagnostou$^{  1}$,
K.J.\thinspace Anderson$^{  9}$,
S.\thinspace Asai$^{ 23}$,
D.\thinspace Axen$^{ 27}$,
I.\thinspace Bailey$^{ 26}$,
E.\thinspace Barberio$^{  8,   p}$
% ...
}\end{center}\bigskip
\bigskip
$^{  1}$School of Physics and Astronomy, University of Birmingham,
Birmingham B15 2TT, UK
\newline
$^{  2}$Dipartimento di Fisica dell' Universit\`a di Bologna and INFN,
I-40126 Bologna, Italy
\newline
$^{  3}$Physikalisches Institut, Universit\"at Bonn,
D-53115 Bonn, Germany
\newline
% ...
$^{  5}$Cavendish Laboratory, Cambridge CB3 0HE, UK
\newline
% ...
$^{  8}$CERN, European Organisation for Nuclear Research,
CH-1211 Geneva 23, Switzerland
\newline
$^{  9}$Enrico Fermi Institute and Department of Physics,
University of Chicago, Chicago IL 60637, USA
\newline
% ...
$^{ 22}$Department of Physics and Astronomy, Tel Aviv University,
Tel Aviv 69978, Israel
\newline
$^{ 23}$International Centre for Elementary Particle Physics and
Department of Physics, University of Tokyo, Tokyo 113-0033, and
Kobe University, Kobe 657-8501, Japan
\newline
% ...
$^{ 26}$University of Victoria, Department of Physics, P O Box 3055,
Victoria BC V8W 3P6, Canada
\newline
$^{ 27}$University of British Columbia, Department of Physics,
Vancouver BC V6T 1Z1, Canada
\newline
% ...
$^{  p}$ now at The University of Melbourne, Victoria, Australia
\newline
% ...
$^{  y}$ now at CERN
\newline
% ...
$^{  *}$ Deceased

\section{Introduction}

The measurement of the mass of the W boson (\mw) is one of the principal
goals of the physics programme undertaken with the LEP \epem\ collider at
CERN. Within the Standard Model of electroweak interactions, the W mass
can be inferred indirectly from precision measurements of electroweak
observables, in particular from $\epem\rightarrow\zb$
events at centre-of-mass energies (\rs) close to the peak of the
Z resonance (around 91\,GeV), studied extensively
at LEP1 and SLD \cite{smew}. These measurements currently give a prediction
for \mw\ with an uncertainty of 32\,MeV, or 23\,MeV if the measurement
of the mass of the top quark from the Tevatron \cite{mtop} is also taken 
into account. Direct
measurements of the W mass with a similar precision are therefore of great
interest, both to test the consistency of the Standard Model and better to
constrain its parameters (for example the mass of the so-far unobserved
Higgs boson), and to look for deviations signalling the possible presence
of new physics beyond the Standard Model. Such measurements became possible
at LEP once the centre-of-mass energy was raised above 160\,GeV in 1996,
allowing the production of pairs of W bosons in the reaction 
$\epem\rightarrow\ww$. Measurements of the width of 
the W boson (\gw) can also be carried out at LEP, providing a further 
test of the consistency of the Standard Model.

This paper presents the final OPAL measurement of the mass and width of
the W boson, using direct reconstruction of the two boson masses in
$\epem\rightarrow\ww\rightarrow\qqlv$ and \qqqq\ events recorded at 
\epem\ collision energies between 170\,GeV and 209\,GeV. The result for \mw\
is combined
with a measurement using direct reconstruction in the \lvlv\ final state 
\cite{mwlvlv} and a measurement from the dependence of the WW production
cross-section on \mw\ at $\sqrt{s}\approx 161$\,GeV \cite{mwthres}.
This paper supersedes our previous results \cite{wmass172,wmass183,wmass189}
obtained from the data with $\rs=$170--189\,GeV.

Three methods are used in this paper to extract \mw, all based on similar
kinematic fits to the reconstructed jets and leptons in each event.
The principal method, the convolution fit, is based on an event-by-event 
convolution of a 
resolution function, describing the consistency of the event kinematics 
with various W boson mass hypotheses, with a Breit-Wigner
physics function dependent on the assumed true W boson mass and width.
The convolution fit is used to obtain the central results of this paper,
but is complemented by two other fit methods of slightly lower statistical 
precision: a reweighting fit based on fitting 
Monte Carlo template
distributions with varying assumed W mass and width to the reconstructed
data distributions, and a simple analytic Breit-Wigner fit to the 
distribution of reconstructed W boson masses in the data. Complete analyses,
including systematic uncertainties, have been performed for all three
methods, providing valuable cross-checks of all stages of the analysis
procedure. The convolution and reweighting fits also measure the W width;
the convolution fit is again used for the central results, and the reweighting
fit provides a cross-check including all systematic uncertainties.
The Breit-Wigner fit does not measure the W width, but an
additional independent convolution-based method is used to provide a 
second statistical cross-check in the \qqlv\ channel.

The dominant systematic error in the \qqqq\ channel comes from possible
final-state interactions (colour reconnection and Bose-Einstein correlations)
between the decay products of the two hadronically
decaying W bosons. According to present phenomenological models, these
interactions mainly affect soft particles, and the uncertainties can 
be reduced by removing or deweighting soft particles when estimating the
directions of jets. Such a method is used for the \qqqq\ channel measurements 
of \mw\ from all three fit methods in this paper. 
Conversely, the effect of final-state interactions
can be enhanced by giving increased weight to soft particles, and this is
used to place constraints on possible colour reconnection effects.

This paper is organised as follows. The OPAL detector, data and Monte Carlo
samples are introduced in Section~\ref{s:dmc}, followed by a brief description
of the event selection in Section~\ref{s:evtsel}. Elements of the event
reconstruction and kinematic fitting common to all three analysis methods
are discussed in Section~\ref{s:reckin}, followed by a detailed description
of the individual convolution, reweighting and Breit-Wigner fits in 
Sections~\ref{s:cvfit}--\ref{s:bwfit}. Systematic uncertainties, which are
largely common to all three methods, are described in Section~\ref{s:syst}.
Finally the results are summarised in Sections~\ref{s:results} 
and~\ref{s:conc}.

\section{Data and Monte Carlo samples}\label{s:dmc}

A detailed description of the OPAL detector can be found elsewhere 
\cite{opaldet}. Tracking of charged particles was performed by a central
detector, enclosed in a solenoid which provided a uniform axial magnetic field
of 0.435\,T.  The central detector consisted of a two-layer silicon 
microvertex detector, a high precision vertex chamber with both axial
and stereo wire layers, a large volume jet chamber providing both tracking
and ionisation energy loss information,
and additional chambers to measure the $z$ coordinate of tracks as they
left the central detector.\footnote{A right handed coordinate system is used,
with positive $z$ along the $\rm e^-$ beam direction and $x$ pointing 
towards the centre of the LEP ring. The polar and azimuthal angles are denoted
by $\theta$ and $\phi$, and the origin is taken to be the centre of the
detector.} Together these detectors provided tracking coverage for polar angles
$|\cos\theta|<0.96$, with a typical transverse momentum ($p_{\rm T}$) 
resolution\footnote{The convention $c=1$ is used throughout this paper.} of 
$\sigma_{p_{\rm T}}/p_{\rm T}=\sqrt{(0.02)^2+(0.0015p_T)^2}$ 
with $p_{\rm T}$ measured in GeV.
The solenoid coil was surrounded by a time-of-flight counter array and 
a barrel lead-glass electromagnetic calorimeter with a presampler. Including
also the endcap electromagnetic calorimeters, the lead-glass blocks covered
the range $|\cos\theta|<0.98$ with a granularity of about $2.3^\circ$ in both
$\theta$ and $\phi$.
Outside the electromagnetic calorimetry, 
the magnet return yoke was instrumented with
streamer tubes to form a hadronic calorimeter, with angular coverage in the
range $|\cos\theta|<0.91$ and a granularity of about $5^\circ$ in $\theta$
and $7.5^\circ$ in $\phi$. The region $0.91<|\cos\theta|<0.99$ was instrumented
with an additional pole-tip hadronic calorimeter using multi-wire chambers,
having a granularity of about $4^\circ$ in $\theta$ and $11^\circ$ in $\phi$.
The detector was completed
with muon detectors outside the magnet return yoke. These were composed
of drift chambers in the barrel region and limited streamer tubes in the
endcaps, and together covered 93\,\% of the full solid angle.
The integrated luminosity was evaluated using small
angle Bhabha scattering events observed in the forward calorimeters 
\cite{opalff}. 

The data used for this analysis were taken at centre-of-mass energies between
170\,GeV and 209\,GeV during the LEP2 running period from 1996 to 2000, and
correspond to a total integrated luminosity of about 689\,\ipb. 
In the year 2000, LEP was operated in a mode where the beam energy was
increased in $\sim 0.5$\,GeV steps during data taking several times in 
each collider fill. Data taken during these `miniramps' (approximately 1\,\%
of the total year 2000 data sample) are excluded from the analysis as the
beam energy is not precisely known.
A detailed breakdown of the energy ranges and integrated
luminosities in each year of data taking is given in Table~\ref{t:evtsel}.
In addition, $\epem\rightarrow\zb$ events recorded at 
$\sqrt{s}\approx 91$\,GeV were used to calibrate the leptonic and hadronic
energy scales and to study the modelling of the detector response by the
Monte Carlo simulation. These events were recorded during dedicated runs
at the beginning of each year, and also at intervals later in the 
data-taking periods to monitor the stability of the detector performance
with time. They amount to a total integrated luminosity of about 13\,\ipb,
corresponding for example to about 400\,000 hadronic \zb\ decays.

\begin{table}
\centering

\begin{tabular}{cccc|cccccccc}
Year & \rs\ range & $\mean{\rs}$ & $\int L$\,dt & \multicolumn{2}{c}{\qqev} & 
\multicolumn{2}{c}{\qqmv} & \multicolumn{2}{c}{\qqtv} & 
\multicolumn{2}{c}{\qqqq} \\
 & (GeV) & (GeV) & (pb$^{-1}$) 
& obs. & exp. & obs. & exp. & obs. & exp. & obs. & exp. \\
\hline
1996 & 170--173 & 172.1 & 10.4 & 
22 & 20 & 15 & 19 & 13 & 10 & 60 & 58 \\
1997 & 181--184 & 182.7 & 57.4 & 
134 & 122 & 117 & 124 & 118 & 124 & 437 & 446 \\
1998 & 188--189 & 188.6 & 183.1 &
388 & 413 & 422 & 417 & 444 & 425 & 1551 & 1511 \\
1999 & 192--202 & 197.4 & 218.5 & 
524 & 512 & 489 & 518 & 559 & 526 & 1924 & 1891 \\
2000 & 200--209 & 206.0 & 219.6 & 
506 & 524 & 530 & 525 & 555 & 543 & 1921 & 1925 \\
\hline
Total & 170--209 & 196.2 & 688.9 & 
1574 & 1591 & 1573 & 1603 & 1689 & 1628 & 5893 & 5831 \\
\hline
\multicolumn{4}{l|}{Estimated selection efficiency (\%)} & &85 & &89 & &68 & &86 \\
\multicolumn{4}{l|}{Estimated purity (\%)} & &92 & &92 & &73 & &79 \\
\end{tabular}
\caption{\label{t:evtsel}Observed and expected numbers of candidate
WW events, together with the collision energy range, mean energy and 
integrated luminosity, in each year of 
data taking. The efficiencies and purities of the event selections,
estimated from Monte Carlo events, are also given.}
\end{table}

Large samples of Monte Carlo simulated events have been generated to 
optimise and calibrate the W mass and width analysis methods, and to 
study systematic uncertainties. The relevant contributions to the
$\epem\rightarrow\qqlv$ and \qqqq\ topologies studied in this paper
can be divided into four-fermion and two-fermion processes \cite{smwidth}.
As defined here, four-fermion
final states ($\epem\rightarrow 4\rm f$) include contributions from
both $\epem\rightarrow\ww\rightarrow 4\rm f$ and 
$\epem\rightarrow\zz\rightarrow 4\rm f$, but exclude multi-peripheral
diagrams resulting from two-photon interactions, which have a negligible
probability of being selected by the analysis requirements and are not
considered further.
Most four-fermion final states were simulated using the {\sc KoralW} 1.42
program \cite{koralw142}, which uses matrix elements calculated
with grc4f 2.0 \cite{grc4f}. These samples were split into two parts, 
corresponding
to four-fermion final states which could have been produced from diagrams
involving at least one W boson (referred to collectively as WW events
below), and others (referred to as \zz\ events, but including some
diagrams not involving two Z bosons). Most WW events
were generated with $\mw=80.33$\,GeV and $\gw=2.09$\,GeV, but samples
with other W masses and widths were also produced in order 
to calibrate and test the fitting procedures. 
The running width scheme for the Breit-Wigner distribution as implemented
in {\sc KoralW} was used throughout. Four-fermion background from the process
$\epem\rightarrow\epem\qqbar$ (included in the \zz\ sample)
was simulated using grc4f.
The only important two-fermion background process is
$\epem\rightarrow\zg\rightarrow\qqbar$, generated using 
KK2f 4.13 \cite{kk2f}, with {\sc Pythia} 6.125
\cite{pythia} as an alternative. 

Hadronisation of final states involving quarks was performed using 
the {\sc Jetset} 7.4 model \cite{jetset}, with parameters tuned by OPAL
to describe global event shape and particle production data at 
the Z resonance \cite{opjetset}.
This hadronisation model and parameter set is denoted by JT.
To study systematic uncertainties related to hadronisation, the same two- and 
four-fermion events have been hadronised with various alternative 
hadronisation models and parameter sets: {\sc Jetset} 7.4 with an
earlier OPAL-tuned parameter set 
based primarily on event shapes \cite{oldjetset} (denoted
JT$'$),
{\sc Ariadne} 4.08 \cite{ariadne} with parameters tuned to ALEPH data 
\cite{alephar} (denoted by AR), {\sc Ariadne} 4.11 (AR$'$) and {\sc Herwig} 6.2 \cite{herwig} (HW),
both with parameters tuned to OPAL data. The possible effects of final-state 
interactions in 
$\epem\rightarrow\ww\rightarrow\qqqq$ events have been studied using 
colour reconnection
models implemented in {\sc Pythia}, {\sc Ariadne} and {\sc Herwig}, and
the LUBOEI Bose-Einstein correlation model \cite{luboei} implemented in 
{\sc Pythia}, as discussed in Section~\ref{s:fsi}. The effects of so-called 
$O(\alpha)$ photon radiation have been studied using the {\sc KandY} 
generator scheme \cite{kandy}, which uses {\sc YFSWW3} \cite{yfsww}
and {\sc KoralW} 1.51 \cite{kandy}
running concurrently, as discussed in detail in Section~\ref{s:oalpha}.

All Monte Carlo samples
have been passed through a complete simulation of the OPAL detector 
\cite{gopal} and the same reconstruction and analysis algorithms as the
real data. Small corrections were applied to the reconstructed jet and
lepton four-vectors in Monte Carlo events better to model the energy
scales and resolutions seen in data, as discussed in detail in
Section~\ref{s:detsys}.

\section{Event selection}\label{s:evtsel}

The selections of $\ww\rightarrow\qqlv$ and $\ww\rightarrow\qqqq$ events
are based on multivariate relative likelihood discriminants, and are
discussed in detail in \cite{wwxsec189}. Events selected by the
$\ww\rightarrow\lvlv$ selection of \cite{wwxsec189} are rejected, and
events selected as both \qqlv\ and \qqqq\ candidates are retained only for
the \qqlv\ analysis.
The sets of reference histograms used
in the selections have been extended to maintain optimal performance for the
highest energy LEP2 running. 

Semileptonic $\ww\rightarrow\qqlv$ decays comprise 44\,\% of the
total WW cross-section, and are selected using separate likelihood
discriminants for the \qqev, \qqmv\ and \qqtv\ channels. These events
are characterised by two well-separated hadronic jets, large missing momentum
due to the escaping neutrino from the leptonic W decay, and in the case
of \qqev\ and \qqmv\ decays, an isolated high-momentum charged lepton. In
$\ww\rightarrow\qqtv$ events, the $\tau$-lepton is identified as an isolated
low multiplicity jet, typically containing one or three tracks. A small
number of `trackless-lepton' \qqev\ and \qqmv\ events are also selected, where
the lepton is identified based on calorimeter and muon chamber
information only, without
an associated track. These events make up 2.5\,\% of the \qqev\ and
4.7\,\% of the \qqmv\ samples. Hadronic
$\ww\rightarrow\qqqq$ decays comprise 46\,\% of the total WW cross-section,
and are characterised by four energetic hadronic jets and little or
no missing energy.  The dominant background results from 
$\epem\rightarrow\zg\rightarrow\qqbar$ events giving a four-jet topology
($\qqbar\rightarrow\qqqq$ or $\rm q\bar{q}gg$),
and this is largely rejected using an event weight based on 
the $O(\alpha_s^2)$ QCD matrix element for this background process.

The number of events selected in each of the channels and data-taking years
is given in Table~\ref{t:evtsel}, together with the expectation from
the Monte Carlo simulation with the WW production cross-section
scaled to the prediction of {\sc KandY} (which is more accurate than 
that of {\sc KoralW}). The average selection 
efficiency and purity of each channel in the desired WW signal topology are
also given, estimated from Monte Carlo events and averaged over all 
centre-of-mass energies. The dominant backgrounds are
events misclassified between the $\qqev/\qqmv$ and \qqtv\ channels
in the \qqlv\ selection, and $\epem\rightarrow\zg\rightarrow\qqbar$ events
giving a four-jet topology in the \qqqq\ channel. 
Combining all three \qqlv\ sub-channels, and including events mis-classified
between them, 87\,\% of \qqlv\ events are selected for the mass and width
analyses. However, not all selected events
are actually used by each analysis---some poorly reconstructed events are 
removed by analysis-specific cuts as discussed below.

\section{W boson reconstruction and kinematic fitting}\label{s:reckin}

All three analysis methods use similar event reconstruction and kinematic
fit techniques to determine the W mass on an event by event basis. In
$\ww\rightarrow\qqlv$ events, the procedure begins by removing the 
tracks, electromagnetic
and hadronic calorimeter clusters corresponding to the lepton identified by 
the event selection. A matching algorithm is then applied to tracks and
calorimeter clusters, and the cluster energies are adjusted both to compensate
for the expected energy sharing between the electromagnetic and hadronic
calorimeters, and to account for the expected energy deposits from any 
associated tracks. This procedure has been optimised to obtain
the best possible jet energy resolution on $\zb\rightarrow\qqbar$ events
at $\rs\approx 91$\,GeV, where use of the hadronic as well as the 
electromagnetic
calorimeter information improves the energy resolution by about 10\,\%.
The reconstructed objects (referred to hereafter as particles)
are then grouped into
two jets using the Durham jet-finding algorithm \cite{durjet}.
 Estimates of the jet energies, directions and
masses are derived from the four-momentum sum of all the tracks and 
corrected calorimeter clusters
assigned to the jet, assigning tracks the pion mass and 
clusters zero mass. Corresponding error matrices are also assigned to the
reconstructed jet energies and directions, based on studies of jet resolution 
in Monte Carlo.

In \qqev\ events, the electron energy is reconstructed from the energy
of the associated electromagnetic calorimeter cluster, and the direction is 
taken from
that of the associated track (except in trackless \qqev\ events, where
both the energy and direction are taken from the calorimeter cluster). 
In \qqmv\ events, the track is used for both
the muon energy and direction estimates. In both cases, 
calorimeter clusters which are not associated to the lepton, but which 
are close to the lepton track and consistent with originating
from final-state radiation, are added into the lepton energy
estimate. In \qqtv\ events, the $\tau$ energy cannot be reconstructed due to
the undetected neutrino(s) produced in its leptonic or hadronic decay. This
means that only the hadronic $\rm W\rightarrow\qqbar$ decay carries usable 
information about the W mass, and the $\tau$ energy and direction are
not reconstructed; this is also the case for trackless \qqmv\ events.
However, a complication can arise in the case of 
hadronic $\tau$ decays if the $\tau$ decay products are incorrectly identified
and some of them mistakenly 
included in the reconstruction of the \qqbar\ system.
The mass information in such events can sometimes be recovered by using 
an alternative $\tau$ reconstruction, forcing the whole event to a three-jet
topology and assuming the $\tau$ to be the jet with lowest invariant mass.
A multivariate procedure based on angular and momentum variables is therefore
used in hadronic $\tau$ decays to decide between the two alternatively
reconstructed topologies.

In $\ww\rightarrow\qqqq$ events, the initial reconstruction used in
the event selection is made
by grouping all tracks and clusters into four jets using the Durham 
algorithm, with double-counting corrected as discussed above. However, a
hard gluon is radiated from one of the quarks in a 
significant fraction of \qqqq\ events, and the mass resolution for such events
can be improved by reconstructing them with five jets \cite{wmass189}.
The convolution and reweighting fits treat all \qqqq\ events in this way,
whilst the Breit-Wigner fit reconstructs the event as four or five jets
depending on the value of $y_{45}$, the value of the Durham jet resolution 
parameter at which the five- to four-jet transition occurs. In all cases,
the jets can be assigned to the two W bosons in several possible ways,
leading to combinatorial background where the wrong assignment has been 
chosen---this is dealt with in different ways by the different analysis
methods as discussed in detail below.

The invariant masses of the two W bosons in the event could be determined
directly from the momenta of the reconstructed jets and leptons, but the
resolution would be severely limited by the relatively poor jet energy
resolution of $\sigma_{E_{\rm jet}}/E_{\rm jet}\approx 12$\,\% for 
well-contained light-flavour jets. For events without significant initial-state
radiation, the W mass resolution can be significantly
improved by using a kinematic fit imposing the four constraints that
the total energy must be equal to the LEP centre-of-mass energy and that
the three
components of the total momentum must be zero (referred to as the 4C fit).
Since the uncertainty 
on the two reconstructed W boson masses is typically still larger than
the intrinsic W boson width of around 2\,GeV, the resolution can be further
improved by constraining the two masses to a common value (the 5C fit). 
The 4C and 5C fits are used in various
ways by the three analysis methods. In the \qqev\ and \qqmv\ channels three 
of the constraints are effectively absorbed by the unmeasured neutrino,
and in the \qqtv\ channel an effective one-constraint fit is performed to the
hadronic part of the event only. In all kinematic fits, the velocity 
of the jet $\beta=p_{\rm jet}/E_{\rm jet}$ is kept fixed as the jet energy 
$E_{\rm jet}$ is varied, which results in the jet momentum
$p_{\rm jet}$ and mass, $m_{\rm jet}=\sqrt{E^2_{\rm jet}-p^2_{\rm jet}}$,
also varying. This procedure is found to give
results which are about 1\,\% more precise than the fixed $m_{\rm jet}$ 
approach used previously \cite{wmass189}.

The dominant systematic error on the measurement of the W mass and width
in the \qqqq\ channel comes from possible final-state interactions between the
decay products of the two W bosons. According to 
phenomenological models, these interactions mainly affect low momentum 
particles produced far from
the cores of the jets. The uncertainties due to final-state interactions
can therefore be reduced by deweighting such particles when 
calculating the jet four-momenta, for example by removing all particles
with momentum $p$ below a certain cut, weighting particles according
to their momentum or only using particles whose directions lie close to
the jet axis.

Such an approach is used for the \qqqq\ channel W mass measurement 
in this paper. The
jet energy and mass are calculated using the original Durham jet definition,
but the jet direction is taken instead
from the sum of the momenta of all particles
assigned to the jet which have $p>2.5\,$GeV. This cut strongly reduces
the systematic uncertainties due to final-state interactions, at the expense
of some loss of statistical precision due to the reduction in jet angular 
resolution. This value of the cut was found to be optimal given the expected
statistical error of the OPAL analysis. In around 4\,\% of jets, no
particles have momenta above 2.5\,GeV, in which case the original jet 
direction is used. For comparison, the \qqqq\ analysis results are also
given using the unmodified Durham jet direction reconstruction (referred
to as $J_0$), though this 
value is not used in the final result.
In the \qqlv\ channel and for the W width analysis, the unmodified $J_0$ 
Durham jet reconstruction is always used.

The sensitivity of the \qqqq\ W mass analysis to final-state interactions
can also be increased, by using a jet direction reconstruction
giving higher weight to soft particles. In the convolution analysis, this is 
done by using a second modified reconstruction method, 
where the jet direction is calculated from the vector sum of the momenta 
of all particles assigned to the jet, each one weighted by 
$p^\kappa$, with $\kappa=-0.5$. The difference between the W mass calculated
using this algorithm (referred to as \syavi) and the algorithm with 
$p>2.5$\,GeV (referred to as \syaiii)
is sensitive to the presence of final-state interactions, and is used
to set a limit on their possible strength within specific models.
Using the same method, but with
positive values of $\kappa$, reduces the sensitivity of the analysis to
final-state interactions, as does using a cone-based direction reconstruction
where only particles within an angle $R$ of the original jet axis are used
to calculate an updated jet direction. Results from these algorithms are
also given in Section~\ref{s:cvres} for comparison purposes.

\section{The convolution fit}\label{s:cvfit}

The convolution fit is based on the event-by-event convolution of
a resolution function $R_i(\mta,\mtb)$ for event $i$ with a physics function
$P(\mta,\mtb |\mw,\gw,\rs)$. The latter represents the expected distribution 
of true event-by-event W masses
\mta\ and \mtb\ given the true W mass and width, and the 
centre-of-mass energy. The resolution function gives the relative probability
that a given observed event configuration could have arisen from an event
with true masses \mta\ and \mtb, and is calculated in different ways for
the \qqlv\ and \qqqq\ channels. The physics function is the same for both
channels, and is given by
\begin{eqnarray}\label{e:physfn}\label{e:rad}
\lefteqn{P(\mta,\mtb\mid\mw,\gw,\rs) =} \\ & & 
a_0 \left[ \cdot 
B(\mta\mid\mw,\gw)\cdot B(\mtb\mid\mw,\gw)\cdot 
S(\mta,\mtb\mid\rsp) \right] \nonumber \otimes I(\rs,\rsp)\,,
\end{eqnarray}
where $a_0$ normalises the integral of $P$ over the (\mta,\mtb) plane to
unity and the symbol `$\otimes$' denotes convolution. The unnormalised 
relativistic Breit-Wigner distribution $B$ is given by
\begin{equation}\label{e:bwig}
B(m\mid\mw,\gw) = \frac{m^2}{(m^2-\mw^2)^2+(m^2\gw/\mw)^2}\,,
\end{equation}
and the phase space term $S$, describing the suppression close 
to the kinematic limit $\mta+\mtb=\rsp$, where \rsp\ is the effective 
centre-of-mass energy after initial-state radiation, 
is given by
\begin{equation}
S(\mta,\mtb\mid\rsp)=
\sqrt{\left(s'-(\mta+\mtb)^2\right)\cdot 
\left(s'-(\mta-\mtb)^2\right)}\,.
\end{equation}
The radiator function $I(\rs,\rsp)$ describes the effect of initial-state
radiation causing an event
of centre-of-mass energy \rs\ to have its effective centre-of-mass energy
reduced to \rsp\, and is given by
\begin{equation}
I(\rs,\rsp) = \beta x^{\beta-1}\ \frac{\sigma(\rsp,\mw)}{\sigma(\rs,\mw)}
\end{equation}
where $\sigma(\rs,\mw)$ is the W-pair production cross section for a given
\rs\ and \mw, $x$ is the normalised initial state radiation photon energy
$x=E_\gamma/\rs$, $E_\gamma=(s-s')/2\rs$, and 
$\beta=(2\alpha/\pi)\log((\rs/m_{\rm e})^2-1)$ where $\alpha$ is the
electromagnetic coupling constant and $m_{\rm e}$ the electron mass
\cite{smwidth}.

The signal likelihood for event $i$,  $\lsig_i(\mw,\gw)$, is calculated from
the convolution of the resolution and physics functions:
\begin{equation}\label{e:conv}
\lsig_i(\mw,\gw) = R_i(\mta,\mtb) \otimes P(\mta,\mtb\mid\mw,\gw,\rs_{i})\, .
\end{equation}
Additional terms $\lzz_i$ and $\lzg_i$ are included to account for the 
presence of background from ZZ and \zg\ production. These likelihoods are 
parameterised using Monte Carlo events and
weighted by event-by-event probabilities $\psig_i$, $\pzz_i$ and $\pzg_i$ 
that the
event comes from each of these sources, derived from the event selection
likelihoods. The total likelihood for event $i$ is then given by
\begin{equation}\label{e:levt}
\levt_i(\mw,\gw) = \psig_i \lsig_i(\mw,\gw) + \pzz_i \lzz_i + \pzg_i \lzg_i \,,
\end{equation}
and the likelihood for the whole sample is given simply by the product of the
individual event likelihoods. The convolution integrals in 
Equations~\ref{e:rad} and~\ref{e:conv} are performed
numerically, and evaluated using a grid of 8100 points 
in the part of the $(\mta,\mtb)$ plane satisfying 
$\rm 100\,GeV<\mta+\mtb<\rs$ and $|\mta-\mtb|<50\,$GeV.

Separate fits are performed to extract the W mass and width. For the mass,
\mw\ is varied to maximise the overall likelihood, with \gw\ determined from
\mw\ by the Standard Model relation \cite{smwidth}
\begin{equation}\label{e:smwidth}
\gw = 3G_{\rm F}\,m^3_{\rm W} (1+2\alpha_s/3\pi)/(2\sqrt{2\pi})\,,
\end{equation}
where $G_{\rm F}$ and $\alpha_s$ are the Fermi and strong coupling constants.
The fitted mass is obtained from the maximum of the likelihood curve,
and then corrected for the biases discussed below. For the W width, \mw\
is kept fixed at 80.33\,GeV and only \gw\ is varied. In the
\qqqq\ channel, the fitted mass does not depend on the assumed width
and {\em vice versa\/}, but in the \qqlv\ channel the width has a small
residual dependence on the assumed W mass. This is corrected at the end of
the fit procedure according to the value derived from the mass fit, a 
simultaneous two-dimensional fit of \mw\ and \gw\ not being possible for
computational reasons.

\subsection{The \bqqlv\ convolution fit}\label{s:cvqqlv}

In $\ww\rightarrow\qqlv$ events, the missing neutrino leads to kinematic
fit solutions with likelihoods which are not Gaussian, 
especially if the constraint that
the two W masses are equal is not applied. The convolution fit
provides a natural framework to exploit all available information 
in the non-Gaussian resolution function $R(\mta,\mtb)$. For each event,
this function is mapped out in the $(\mta,\mtb)$ plane by performing many
six-constraint kinematic fits, where in addition to energy-momentum 
conservation, the two W masses are fixed to the
input values \mta\ and \mtb\ rather than being left free to be determined 
in the fit.  Each fit therefore gives only a $\chi^2$ value, which varies
as a function of \mta\ and \mtb\ and expresses the consistency of the
event with the input W mass hypothesis. The minimum $(\chi^2_{\rm min})$
of this $\chi^2$ contour
corresponds to the fitted values of the two W masses, \ma\ and \mb, which 
would have been returned by a standard 4C fit. The resolution function $R$
at each point is derived from the $\chi^2$ contour via the relation
$R=\exp((\chi^2_{\rm min}-\chi^2)/2)$ and normalised so that its integral
is unity over the $(\mta,\mtb)$ plane. 

The kinematic fits in the \qqev\ and \qqmv\ channels are performed
using semi-analytic approximations with two simplifying assumptions, namely 
that the lepton direction is fixed, and that the fitted jet directions are
constrained to lie in the plane defined by their measured values. These
allow the fit to be reduced to a one-dimensional numerical minimisation.
If the event is very badly measured (or is in fact a background event), 
the apparent minimum of the $\chi^2$ contour may lie at the 
edge of the mass grid---this happens in about 5\,\%
of \qqev\ and \qqmv\ candidates. An attempt is made to recover some useful
W mass information from such events by discarding the lepton and refitting
them as $\qqtv$ events (where the lepton information is never used); this is 
also done for trackless-lepton \qqmv\ candidates
which have no useful estimate of the lepton energy and only a poor estimate
of its direction from the muon chambers.

The kinematic fit for \qqtv\ events involves only the hadronic system, the
only variables of interest being the angle between the two jets from
the $\rm W\rightarrow\qqbar$ decay and the sharing of the available beam energy
between them. The resolution function $R$ is mapped out using the
same technique as for \qqev\ and \qqmv\ events. Events with a solution within 
2.5\,GeV of any edge of the mass grid are not considered further; this happens
to about 25\,\% of signal $\ww\rightarrow\qqtv$ events and 58\,\% of background
\qqtv\ candidates.

The non-WW sources of background in the \qqlv\ channel are very small in
all but the \qqtv\ case. Their contributions are accounted for by 
background terms in the likelihood (see Equation~\ref{e:levt}) which 
are parameterised as functions of \ma\ and \mb, the two W masses at the 
$\chi^2$ minimum of the kinematic fit solutions, and \rs. Separate
parameterisations are used for \qqev, \qqmv, \qqtv\ and trackless-lepton
\qqev\ and \qqmv\ candidates, derived from
large samples of background Monte Carlo simulated events.

In Monte Carlo \qqlv\ events, the W mass and width estimates derived from the 
convolution fit differ from the simulated values 
by up to 350\,MeV, due to effects not fully accounted
for in the likelihood function, for example biases in the input jet and 
lepton four-vectors, 
and imperfections in the treatment of initial-state radiation and backgrounds.
These biases are studied by applying the convolution fit to 
large Monte Carlo samples of simulated signal and background events with
various true values of \mw\ from 79.33\,GeV to 81.33\,GeV, and \gw\ from
1.6\,GeV to 2.6\,GeV. For the W mass fit, the biases are
found to depend on \rs\, but not on the true values \mw\ and \gw, and
are parameterised from Monte Carlo as smooth functions of \rs.
In the width fit, the bias on the reconstructed width is found to depend 
slightly on the true width, as well as on the true mass as discussed above.
These biases are again parameterised using Monte Carlo.
The errors returned by the fits are also checked,
by studying pull distributions
obtained from fits to many Monte Carlo subsamples constructed so as to have
the same integrated luminosity as the data in each year. These studies show
that the fits underestimate the statistical error by about 5\,\%, 
reflecting imperfections in the input jet and lepton error matrices.
Corresponding corrections are therefore applied to the statistical
errors determined by the fits. 

After these corrections, the fits give unbiased
results on Monte Carlo samples, but several further small corrections, 
amounting to a total of about 5\,MeV for the W mass and 20\,MeV for the W 
width, are applied to the data results.
These account for effects not present in the Monte Carlo samples used to
calculate the bias corrections, namely additional non-simulated detector 
occupancy, deficiencies in the description of kaon and baryon production in 
the JT hadronisation model, and $O(\alpha)$ photon radiation modelled
by {\sc KandY} but not {\sc KoralW}. These corrections, which are
also applied to the results of the reweighting and Breit-Wigner fits,
are discussed individually in more detail in Section~\ref{s:syst}.

A second convolution-based fit method (referred to as the `CV5' fit)
is used to make an additional 
cross-check of the W width fit result in the \qqlv\ channel. This method
is similar to the fit described above, except that the two input W masses
\mta\ and \mtb\ are set to be equal, making it equivalent to a 5C rather
than a 4C fit, and tracing out the $\chi^2$ probability contour only along
the diagonal $\mta=\mtb$ in the $(\mta,\mtb)$ plane. This reduces the
number of kinematic fits needed per event, allowing the $\chi^2$ values
to be determined using numerical minimisation rather than the fast
analytic approximations used above. In this fit, a single Breit-Wigner
distribution
is used in the physics function analogous to Equation~\ref{e:physfn}, and both
the mass \mw\ and width \gw\ are determined simultaneously. Similar bias 
correction procedures and parameterisations are used as for the standard
convolution fit.

\subsection{The \bqqqq\ convolution fit}

The \qqqq\ channel differs from the \qqlv\ channel in several important 
respects: no prompt neutrinos are produced, leading to better constrained 
kinematics, but the assignment of jets to the two decaying W bosons 
is ambiguous, leading to combinatorial background where the wrong
assignment is made. Non-WW background (particularly from 
$\epem\rightarrow\zg$ events producing four jets) is also much more
important than in \qqlv\ events, contributing 16\,\% of the selected sample.

In a significant fraction of $\ww\rightarrow\qqqq$ events, a hard gluon is
radiated from one of the quarks, and these events are better reconstructed
as five-jet rather than four-jet events. Since the division between four
and five jets is rather arbitrary, the convolution fit reconstructs all
\qqqq\ events with five jets. In \qqqq\ events with no hard
gluon radiation one of the quark jets is split in two by this
procedure, but the two jet
fragments have a high probability to be correctly assigned to the same W
boson. A more serious problem is the combinatorial background---with five
jets there are ten possible assignments of the jets to two W bosons, 
compared with only three in a four-jet topology. This is dealt with in two
ways. Firstly, only 4C fits are used, where the two W boson masses are
not constrained to be equal; many of the incorrect jet assignments
give kinematic fit solutions with two very different masses, in contrast
to the correct solution with two similar masses. Secondly,
energy ordering of the jets is used together with an artificial 
neural network algorithm based on the 4C fit mass differences
to weight each remaining jet assignment combination in the likelihood
fit. 

In more detail, an initial 4C kinematic fit imposing four-momentum 
conservation is applied to the five jets, which are then ordered
according to their fitted energies. The event is also reconstructed in
a four-jet topology using the Durham scheme, resulting in two of the five jets 
being combined, the other three remaining unchanged. The three unchanged jets
are labelled 1--3 such that $E_1>E_2>E_3$, whilst the remaining two jets
in the five-jet topology
are labelled 4 and 5, with $E_4>E_5$, where $E_i$ refers to the fitted
energy of jet $i$ from the initial 4C kinematic fit. These jets can be assigned
to two W bosons in ten different ways, with the combinations numbered
(124,35), (125,34), (12,345) and so on. The combination (123,45)
is not considered further, as it 
has one W boson formed from just the split jets, which is very 
unlikely. For each of the nine remaining combinations $c$, the jet four-vectors
resulting from the 4C fit are combined to calculate the
reconstructed masses \mac\ and \mbc, and the 
mass difference $\delta m_c=\mac-\mbc$. The nine mass 
differences are input to an artificial neural network \cite{jetnet}
with seven outputs, 
corresponding to each of the remaining combinations apart from
(124,35) and (134,25), which are also discarded at this stage, having
little probability of being correct due to the large imbalance in 
the energies assigned to the two W bosons.
The network is trained using a large sample of signal WW Monte Carlo
events to give values close to one at the output corresponding to 
the correct combination, and zero for all other outputs. In each
\qqqq\ event, the seven outputs are normalised to sum to unity,
and all combinations with output $Q_c>0.12$ are retained for the
final likelihood fit. This cut value is found to minimise the statistical
error on the W mass in Monte Carlo events.

The distribution of $Q_c$ for all jet combinations with $Q_c>0.05$
in data is shown in Figure~\ref{f:comb}(a), together
with the expectation from Monte Carlo, broken down into correct and
wrong combinations in WW events, and ZZ and \zg\ background.
The fraction of WW Monte Carlo jet combinations which are 
correct\footnote{A jet combination is considered to be correct if all the 
jets are assigned to the correct W bosons, a jet being correctly assigned
if more than half of its energy results from the decay
products of the associated boson. In about 2\,\% of selected events
no combination is
considered correct, due to more than three jets being assigned to one boson
according to this definition.}
is shown as a function of $Q_c$ in Figure~\ref{f:comb}(b).
The probability for each of the jet combinations to be 
correct before the neural network selection is shown in 
Figure~\ref{f:comb}(c), showing the power of the initial energy 
ordering in already distinguishing the correct combination.
The number of combinations $N_{\rm comb}$ with $Q_c>0.12$ is shown in 
Figure~\ref{f:comb}(d)---typically~3 or~4 combinations are
retained for the final fit. Some discrepancies between data and
Monte Carlo are visible; these are addressed in the systematic
uncertainty studies as discussed in Section~\ref{s:other}.

\epostfig{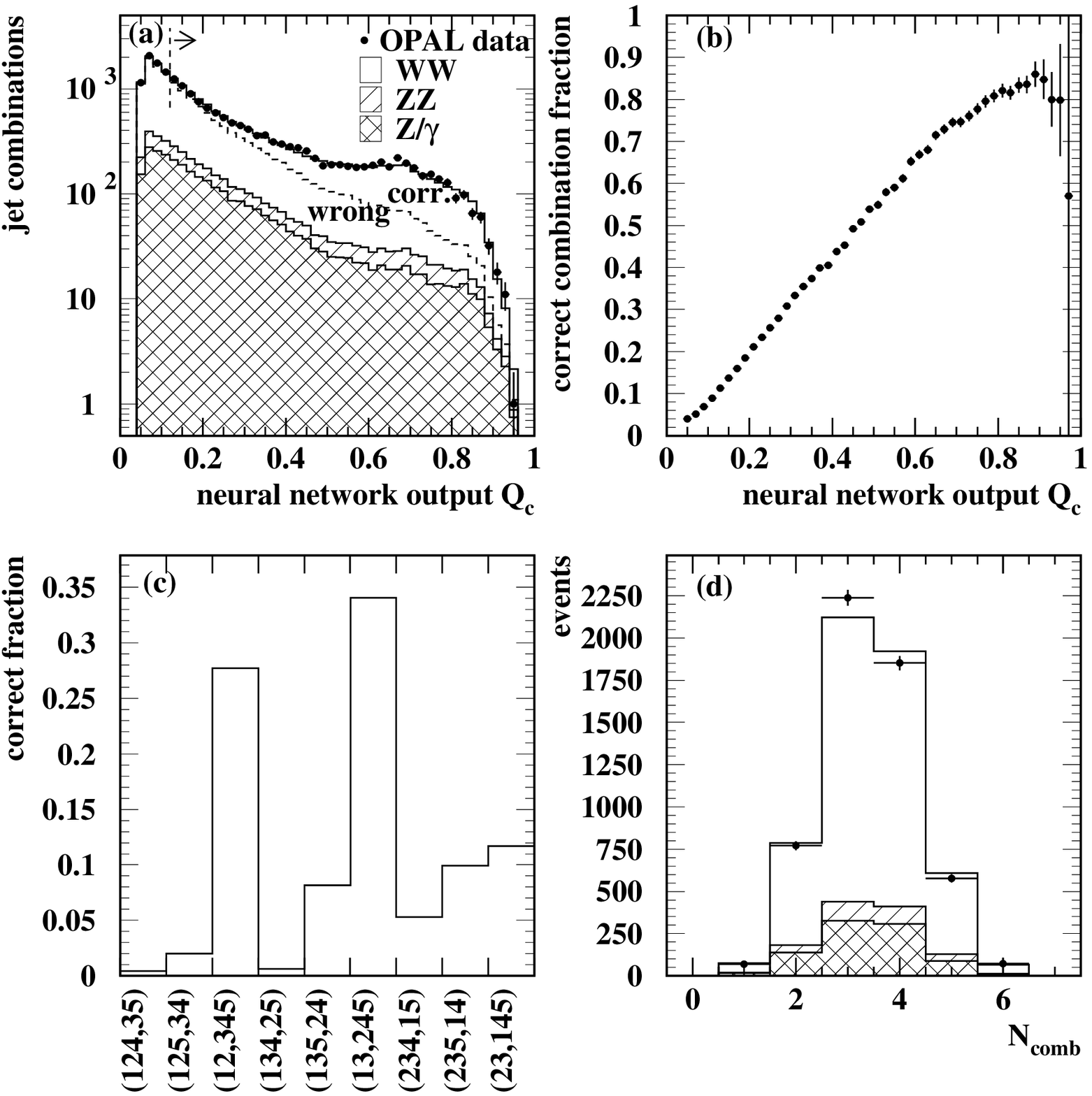}{f:comb}{Details of jet assignment for
the \qqqq\ convolution analysis: (a) distribution of neural network
outputs $Q_c$ for jet combinations in data and Monte Carlo,
showing the correct and wrong combinations in WW events (separated by the
dashed line), the contributions from \zz\ and
\zg\ background events, and the cut used to select combinations for fitting; 
(b) fraction of correct combinations in Monte Carlo WW events as a function of
$Q_c$;
(c) probabilities that each jet combination
is correct, based on energy-ordering before the neural network selection
(note that combinations (124,35) and (134,25) are never used in the fit);
(d) number of accepted combinations $N_{\rm comb}$ 
per event in data and Monte Carlo.}

The resolution function for each retained combination is generated
from the fitted W masses \ma\ and \mb\ returned by the
4C kinematic fit, together with their associated errors and correlation
coefficient. This simple approach, rather than mapping out the full resolution
function using many six-constraint fits with fixed input W masses, 
is adequate due to the 
better-constrained kinematics compared with the \qqlv\ channel.
However, to
model the tails better, a two-dimensional double Gaussian resolution function 
is used, with 
separate core and tail components. The core has a width given by the
event-by-event kinematic fit errors and a weight of 58\,\%, whilst the tail 
component contributes the remaining 42\,\% of the resolution function
and has a width 2.2 times larger than the core.
These associated parameters were derived from studies
of the fit resolution in Monte Carlo simulation.

The signal likelihood function is more complicated than that for \qqlv\ 
events as it must account for the several jet assignment combinations
in each event. This is achieved by treating the W boson masses
\mac\ and \mbc\ for each combination $c$ as independent observables.
For each combination, one pair of masses is described by the
convolution of signal resolution and physics functions, whilst the
others (considered to come from combinatorial background) are each described
by a parameterised function $C(\ma,\mb,\sigma_+,\rs)$, where 
$\sigma_+$ is the error on the sum of the two masses $\ma+\mb$. This function
is obtained from the distributions of combinatorial background combinations
in Monte Carlo events. The 
likelihoods for each of the $N_{\rm comb}$ combinations are then summed, 
weighting each one by its associated neural network output $Q_c$. 
Thus, the signal likelihood for one event is given by
\begin{eqnarray}\label{e:lqqqq}
\lsig_i(\mw,\gw) & = & \sum_{c=1}^{N_{\rm comb}} Q_cd_c  \\
d_c & = & R^c_i(\mta,\mtb)\otimes P(\mta,\mtb\mid\mw,\gw,\rs_i)
\prod_{j=1;j\neq c}^{N_{\rm comb}} C(m^j_1,m^j_2,\sigma^j_+,\rs)\ . \nonumber
\end{eqnarray}
The overall event likelihood is again given by
Equation~\ref{e:levt}. In the case of the \qqqq\ channel, 
the likelihood $\lzz$ for ZZ events
is also given by Equation~\ref{e:lqqqq}, with \mw\ replaced by the known
\mz, and the likelihood $\lzg$ for \zg\ events is given by a similar
expression but with no correct combination, only terms involving
combinatorial background $C(\ma,\mb,\sigma_+,\rs)$. The parameterised
functions $C$  are determined from Monte Carlo separately for each type of
event, and the event type probabilities $\psig$, $\pzz$ and $\pzg$ are 
also parameterised, as linear functions of the event selection likelihood.
 
As for the \qqlv\ fit, bias corrections are applied 
to the raw mass fit results, parameterised as a function of \rs. These
corrections are calculated separately for the fits using the $J_0$ and
modified  jet direction reconstruction methods, and 
are largest (up to 400 MeV) for the \syavi\ jet reconstruction. No significant
dependence of the corrections on the true value of \mw\ is observed.
In the \qqqq\ channel, the width fit bias is also found to be independent
of the true value of \gw, and on the assumed value of \mw. Monte Carlo
subsample tests are also performed, and small corrections to the
fit error estimates of typically 5--10\,\% are derived. Further 
small corrections of up to 9\,MeV
are applied for effects not present in the default Monte Carlo samples,
as discussed in Section~\ref{s:cvqqlv}.

\subsection{Convolution fit results}\label{s:cvres}

The convolution fit is used to analyse the data for each year separately,
and the results are then combined. 
The results and associated statistical uncertainties
are given in Table~\ref{t:mresstat}, for the \qqev,
\qqmv, \qqtv, combined \qqlv\ and \qqqq\ channels.
In the \qqqq\ channel,
the results are given for jet direction reconstruction methods \syaiii\ and
\syavi\, and for comparison also with
the unmodified Durham jet algorithm ($J_0$) as used
in the \qqlv\ channel. The quoted results include all 
corrections made to the fit results as discussed above,
but the averages do not include the effects of systematic uncertainties 
(the final results including all uncertainties are given in 
Section~\ref{s:nonlres}). Table~\ref{t:mresstat} also gives the expected
statistical errors for each channel, evaluated using fits to many 
Monte Carlo subsamples, each constructed to have the same integrated
luminosity and centre-of-mass energy distribution as the data. In all cases, 
the data statistical errors are consistent with the expectations from
Monte Carlo, after taking into account the expected
level of statistical fluctuations. 

\begin{table}
\centering

\begin{tabular}{l|cc|cc|cc}
 & \multicolumn{2}{c|}{Convolution} &
\multicolumn{2}{c|}{Reweighting} & \multicolumn{2}{c}{Breit-Wigner} \\
Channel & Fitted \mw & \sexp & Fitted \mw & \sexp & Fitted \mw & \sexp \\
 & (GeV) & (GeV) & (GeV) & (GeV) & (GeV) & (GeV) \\ \hline
\qqev & $80.511\pm 0.084$ & 0.088 & $80.492\pm 0.088$ & 0.091 & 
    $80.500\pm 0.100$ & 0.104 \\
\qqmv & $80.432\pm 0.090$ & 0.089 & $80.488\pm 0.093$ & 0.091 & 
    $80.523\pm 0.105$ & 0.106 \\
\qqtv & $80.354\pm 0.126$ & 0.125 & $80.289\pm 0.130$ & 0.130 & 
    $80.269\pm 0.135$ & 0.140 \\ \hline 
\qqlv & $80.451\pm 0.056$ & 0.056 & $80.451\pm 0.057$ & 0.058 & 
   $80.457\pm 0.064$ & 0.065 \\ 
\qqqq\ (\syaiii) & $80.353\pm 0.060$& 0.059 & $80.308\pm 0.064$ & 0.066 & 
     $80.286\pm 0.073$ & 0.075 \\ \hline 
\qqqq\ ($J_0$) & $80.394\pm 0.051$ & 0.051 & $80.383\pm 0.056$ & 0.057 & 
     $80.424\pm 0.059$ & 0.060 \\
\qqqq\ (\syavi) & $80.508\pm 0.073$ & 0.073 & --- & ---  & --- &  ---
\end{tabular}
\caption{\label{t:mresstat}W mass results (with statistical errors only)
for each channel and fitting method. The expected statistical errors \sexp\
from Monte Carlo subsample tests are also given. In the \qqqq\ channel,
results are given for the \syaiii\ jet direction reconstruction, the 
$J_0$ reconstruction and the \syavi\ reconstruction giving increased weight
to low-momentum particles (the latter for the convolution fit only).}
\end{table}

Distributions
of the mean of the two W masses reconstructed in each event ($(\ma+\mb)/2$)
are shown for the \qqlv\ fit and sub-channels in 
Figure~\ref{f:qqlvmasscv}, and for the \qqqq\ fit (with jet direction 
reconstruction
\syaiii) in Figure~\ref{f:qqqqmasscv}. In the latter figure, 
reconstructed mass combinations are shown for two ranges of $Q_c$, 
showing the suppression of the 
combinatorial background achieved by the neural network algorithm.
The results obtained in each year of data-taking are shown as the `CV'
points in 
Figure~\ref{f:mresyear}; all the results are consistent with the overall
mean for each channel, and the $\chi^2$ values for the \qqlv\ and \qqqq\ 
(\syaiii) averages are 4.3 and 1.0, each for four degrees of freedom.

\epostfig{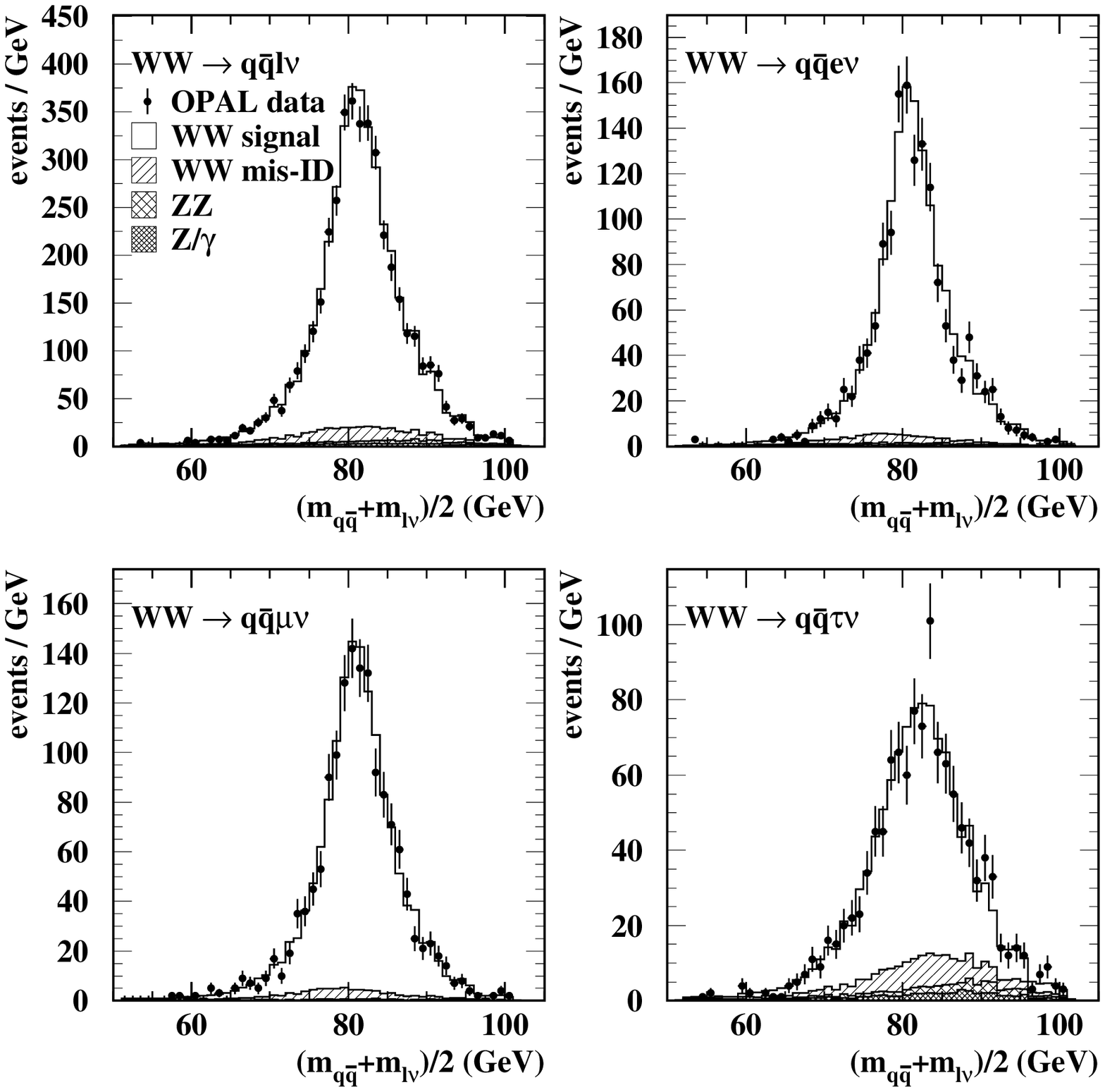}{f:qqlvmasscv}{Reconstructed mean mass 
distributions for $\ww\rightarrow\qqlv$, \qqev, \qqmv\ and \qqtv\ 
candidates fitted using the convolution analysis. The
points with error bars show the data, and the histograms show the 
Monte Carlo expectation (with $\mw=80.415$\,GeV), broken down into 
contributions from signal WW events
with the correct lepton type, WW events with mis-identified leptons 
(`WW mis-ID'), and background from ZZ and \zg\ events.}

\epostfig{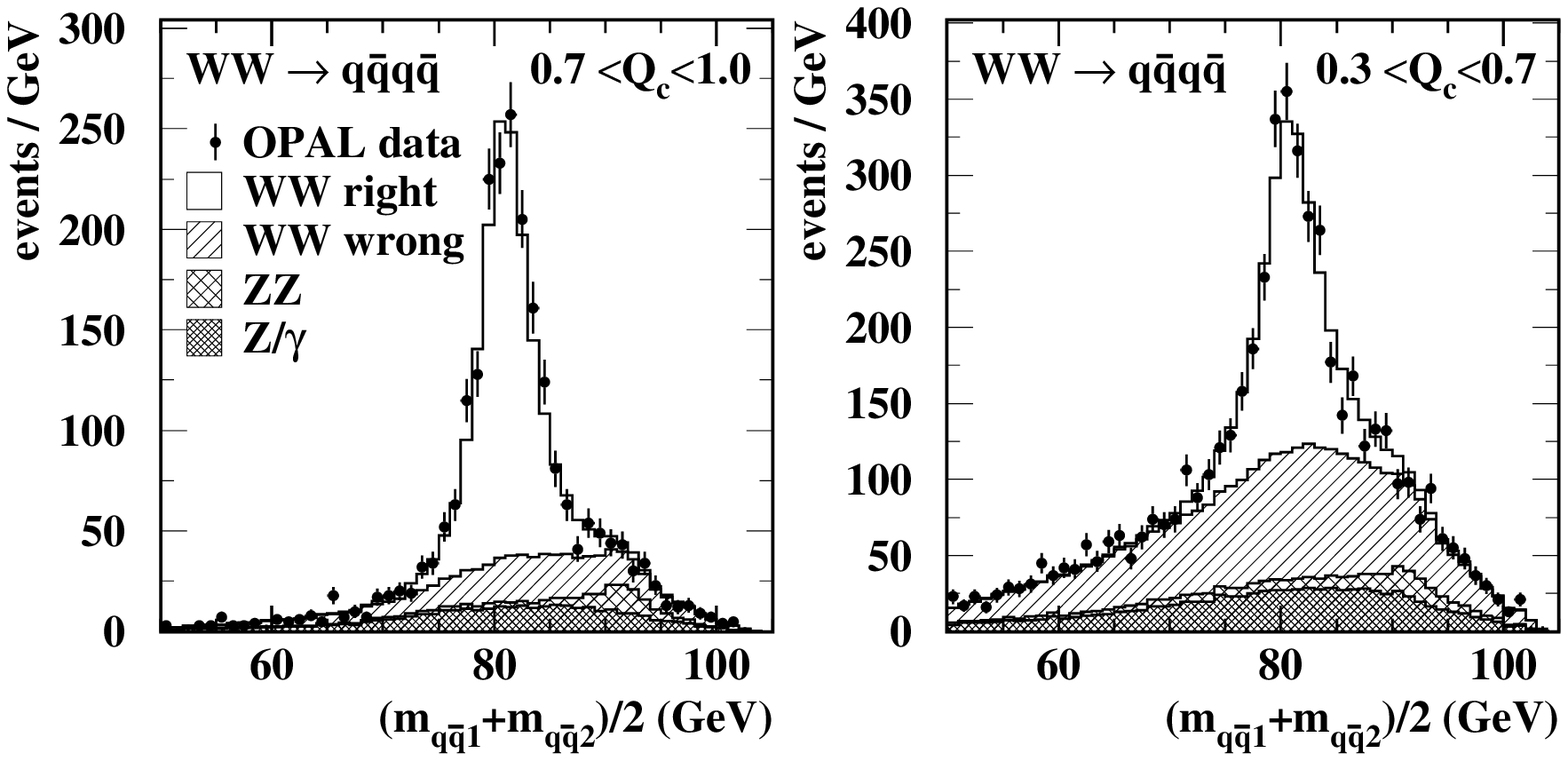}{f:qqqqmasscv}{Reconstructed
mean mass distributions for $\ww\rightarrow\qqqq$ combinations fitted
using the convolution analysis with the  \syaiii\ jet direction 
reconstruction, for two different ranges of correct combination 
probability $Q_c$.
The points with error bars show the data, and the histograms show 
the Monte Carlo expectation (with $\mw=80.415$\,GeV), 
broken down into contributions from correct jet combinations in 
signal WW events, combinatorial background in WW events,
and background from ZZ and \zg\ events.}

\epostfig{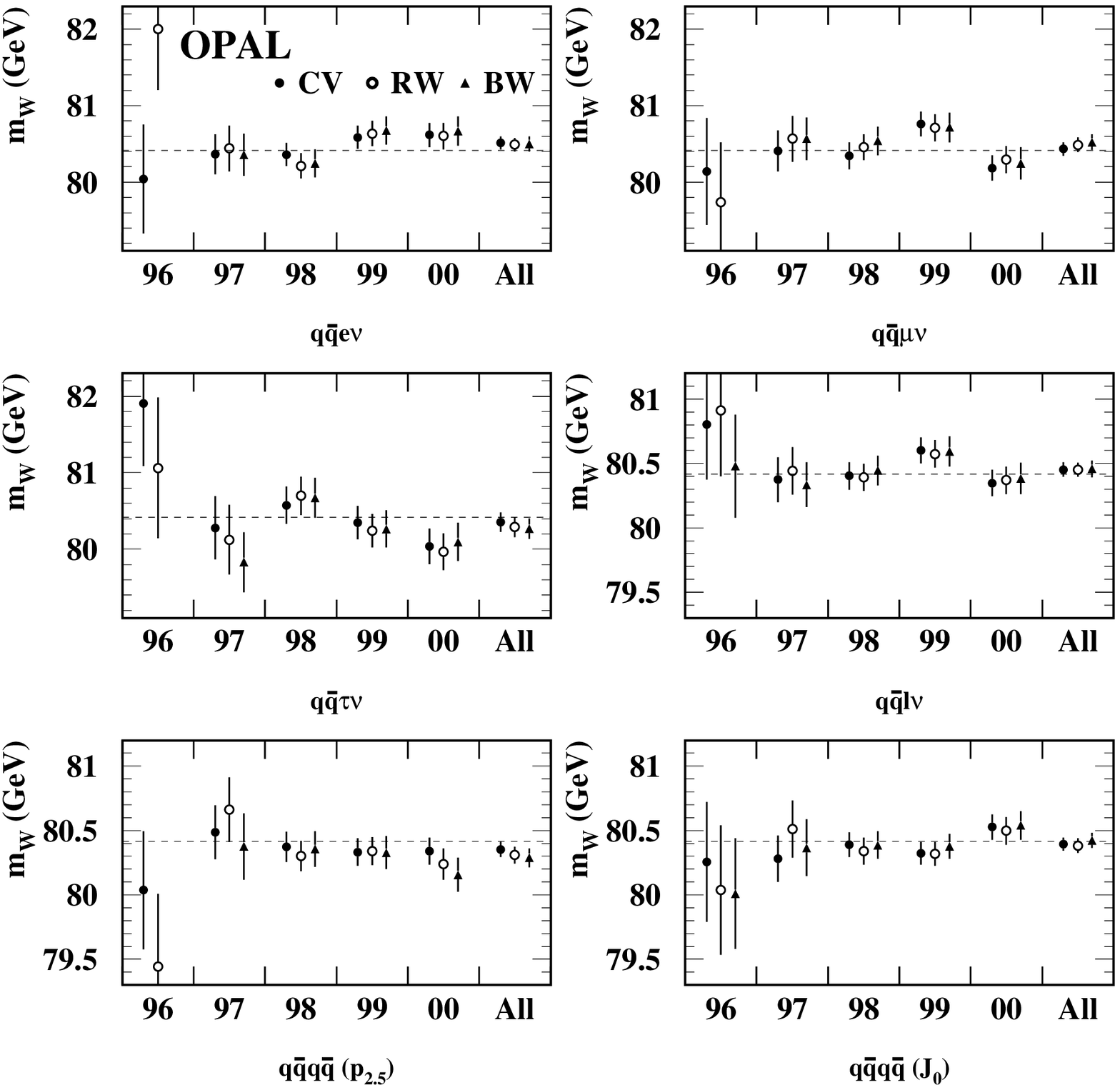}{f:mresyear}{W mass results (with statistical
errors only) for each channel in the convolution (CV), reweighting (RW) and
Breit-Wigner (BW) fitting methods, as a function of
data-taking year, and for all years combined. Results are shown for
the \qqev, \qqmv, \qqtv\ and combined \qqlv\ channels, and for the
\qqqq\ channel with the modified (\syaiii) and  $J_0$ jet
direction reconstruction methods.
The results for the Breit-Wigner \qqlv\ and \qqqq\ fits to 1996 data 
are taken from \cite{wmass172}. The dotted line indicates the central measured 
value of \mw\ from the present paper.}

The corresponding results for the width are shown in Table~\ref{t:wresstat}
and Figure~\ref{f:wresyear}. The 1996 data at $\rs\approx 172$\,GeV are not
used for the width analysis. Again, the statistical uncertainties are 
compatible with expectations, and the individual year results are 
consistent, the $\chi^2$ values being 5.4 and 2.1 for the \qqlv\ and \qqqq\
channels, each for three degrees of freedom. The statistical correlation
between the \qqlv\ mass and width results is estimated using
Monte Carlo subsamples to be $-0.19$, whereas that for the \qqqq\ channel
results is found to be negligible.
Note that the \qqqq\ width analysis is performed using the unmodified $J_0$
jet direction reconstruction as this gives the optimal balance between
statistical and systematic errors from hadronisation and final-state 
interactions, and minimises the total error. The width result from the 
CV5 convolution fit in the \qqlv\ channel is also shown; this fit
also measures the W mass and gives a result of $80.424\pm 0.077$\,GeV, 
consistent with that derived from the standard convolution fit. The statistical
correlation coefficient between the CV5 width and mass fit results is 0.28.

\begin{table}
\centering

\begin{tabular}{l|cc|cc|cc}
 & \multicolumn{2}{c|}{Convolution} &
\multicolumn{2}{c|}{Reweighting} & \multicolumn{2}{c}{CV5 convolution} \\
Channel & Fitted \gw & \sexp & Fitted \gw & \sexp & Fitted \gw & \sexp \\
 & (GeV) & (GeV) & (GeV) & (GeV) & (GeV) & (GeV) \\ \hline
\qqev & $1.696\pm 0.202$ & 0.204 & $2.009\pm 0.200$& 0.202 & 
$1.975\pm 0.230$ & 0.224 \\
\qqmv & $2.181\pm 0.233$ & 0.216 & $2.146\pm 0.224$ & 0.227 & 
$2.138\pm 0.233$ & 0.218 \\
\qqtv & $1.763\pm 0.289$ & 0.289 & $2.089\pm 0.276$ & 0.309 & 
$2.204\pm 0.188$ & 0.241 \\ \hline
\qqlv & $1.926\pm 0.135$ & 0.134 & $2.088\pm 0.131$ & 0.127 & 
$2.103\pm 0.120$ & 0.131 \\  
\qqqq\ ($J_0$) & $2.125\pm 0.111$ & 0.114 & $2.176\pm 0.129$ & 0.134 
& --- & --- 
\end{tabular}
\caption{\label{t:wresstat} W width results (with statistical errors only)
for each channel and fitting method. The expected statistical errors \sexp\ 
from Monte Carlo subsample tests are also given. The Breit-Wigner fit does not
measure the W width.}
\end{table}

\epostfig{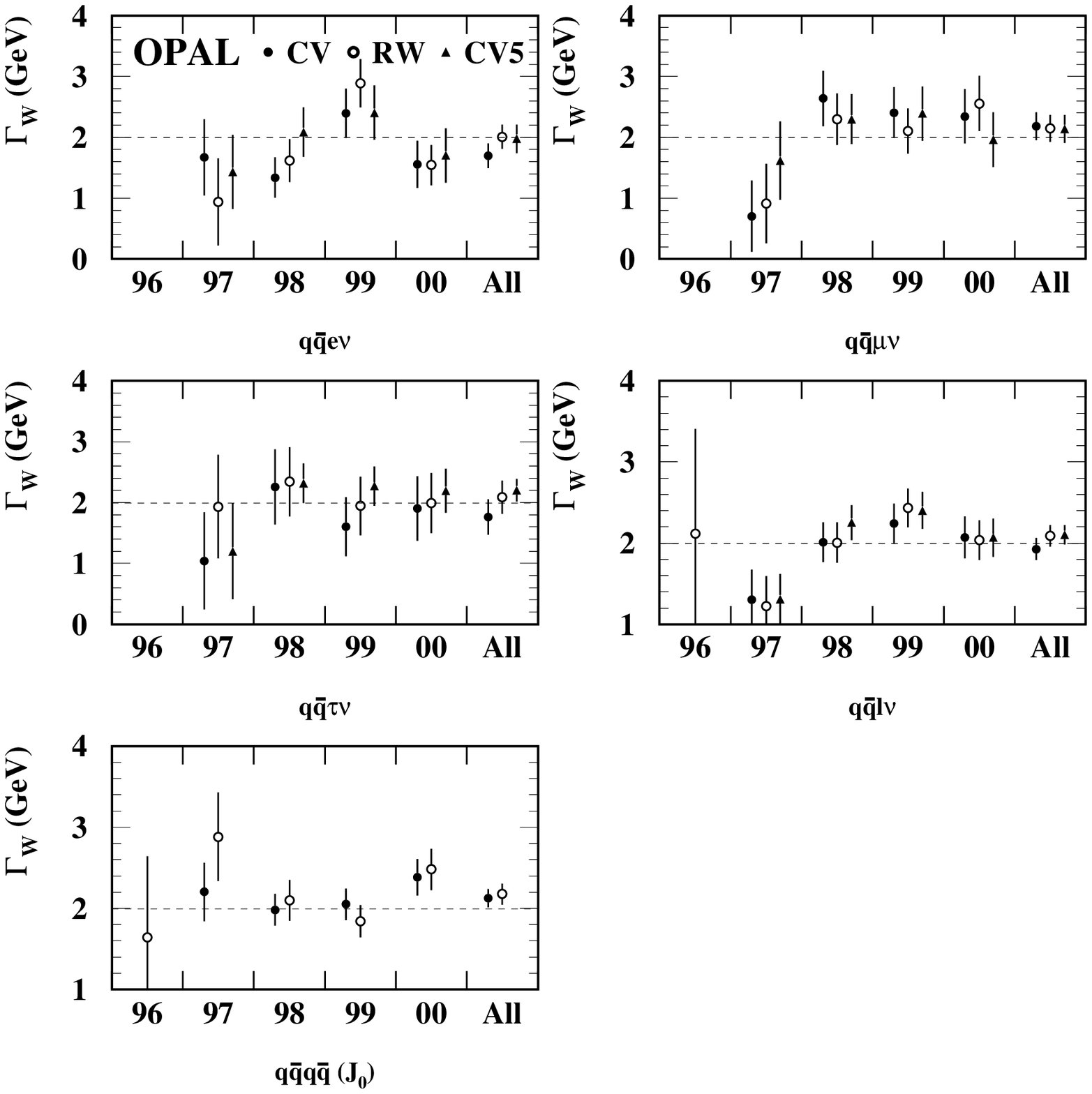}{f:wresyear}{W width results (with statistical
errors only) for each channel in the convolution (CV), reweighting (RW) and
5C convolution (CV5) fitting methods, as a function of
data-taking year, and for all years combined. Results are shown for
the \qqev, \qqmv, \qqtv\ and combined \qqlv\ channels, and for the
\qqqq\ channel with the $J_0$ jet algorithm. Only the
reweighting fit measures the W width using the 1996 data, where results
are shown for the \qqlv\ and \qqqq\ channels only.
The dotted line indicates the central measured
value of \gw\ from the present paper.}

In order to study the evolution of the fitted W mass with changing jet
direction reconstruction, the complete convolution fit has been repeated
fifteen times, using momentum
cuts at 1.0, 1.75, 2.0, 2.5, 3.0 and 4.0\,GeV (2.5\,GeV being used for the 
\qqqq\ 
analysis result in this paper), momentum weights $p^\kappa$ with $\kappa$
values of 0.5, 0.75, 1.0, $-0.5$ and $-0.75$, and cones of half-angle 
$R=$0.3, 0.4, 0.5 and 0.6\,rad. For
each jet direction reconstruction method, the mass difference with respect to 
the
$J_0$ direction reconstruction using all particles associated to the 
jet $\dmjz=m(X)-m(J_0)$ is calculated in both
the \qqlv\ and \qqqq\ channels. The statistical error on $\dmjz$ is also
calculated, using Monte Carlo subsamples to take into account the correlation
due to the different reconstruction methods being applied to the same events.
The results are shown in Figure~\ref{f:delm}.
The \qqlv\ \dmjz\ values are generally 
slightly positive, but the changes in fitted
W mass are consistent with the expected level of statistical fluctuations,
demonstrating that the \qqlv\ W mass results are stable with respect to 
changing the jet direction reconstruction over all methods and 
a wide range of parameter values. The \qqqq\ \dmjz\ results also tend
to be close to zero, with the exception of the result from jet
direction reconstruction \syavi\ (with enhanced sensitivity to low momentum 
particles), which 
is significantly higher ($\dmyz{\syavi}=114\pm 47$\,MeV) than the results 
with all other jet direction
definitions. The result from method \syaxiv\ also shows a high
value of \dmyz{\syaxiv}, although with low significance. Note that 
hadronisation uncertainties are also significant in these comparisons, and 
increase
to around 20\,MeV for the alternative jet direction reconstruction methods,
as discussed in Section~\ref{s:hadrn}.

Sensitivity to final-state interactions in the \qqqq\ channel 
can be maximised by studying the variable 
\begin{equation}
\dmjx=m(X)-m(\syavi)\,,
\end{equation}
the difference in W mass between jet direction
reconstruction methods with reduced and increased sensitivity to these effects.
The largest deviation from zero is seen for 
$\dmyx{\syaiii} = \dmres\pm\dmstat$\,MeV,
where the error is purely statistical, but takes into account correlations
between the different reconstruction methods. The use of the mass differences
to place limits on the effect of final-state interactions, 
specifically colour reconnection, is discussed in Section~\ref{s:fsilim}.

\epostfig{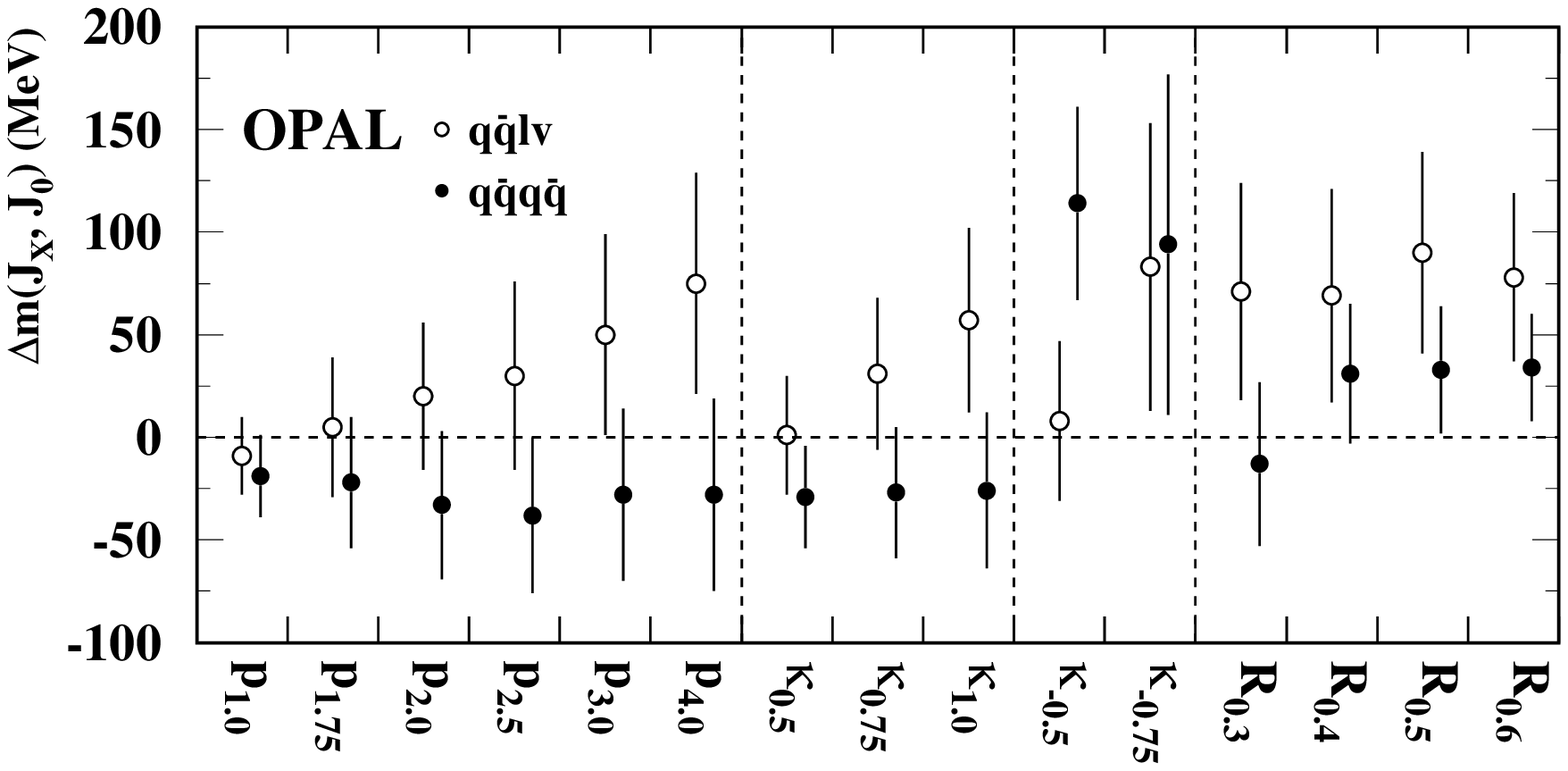}{f:delm}{Differences in \mw\ measured using
various jet direction reconstruction methods $X$ and the $J_0$ 
reconstruction method,
for both \qqlv\ and \qqqq\ data in the convolution fit. The points are 
highly correlated; 
the errors are purely statistical and take into account the correlation
between the result from each alternative direction reconstruction method
and that from $J_0$.}

\section{The reweighting fit}\label{s:rwfit}

The reweighting fit extracts the W mass and width by comparing
reconstructed data distributions with Monte Carlo `template' distributions
with varying \mw\ and \gw. Templates of
arbitrary \mw\ and \gw\ are obtained by reweighting Monte Carlo simulated data
samples containing all signal and background final states,
and a maximum likelihood fit is used to find the 
values of \mw\ and \gw\ that best describe the data.
The reweighting fit is a more sophisticated development of that used
in \cite{wmass189}, the main changes being the use of simultaneous 
reweighting in three (\qqev, \qqmv\ and \qqqq) or two (\qqtv)
reconstructed variables, and an improved procedure
for handling the combinatorial background in the \qqqq\ channel.

In more detail, the likelihood for each event $i$ is given by 
\begin{equation}\label{e:rwlike}
\levt_i(\valp_i|\mw,\gw) = \psig \lsig(\valp_i|\mw,\gw) + 
\pzz \lzz(\valp_i) + \pzg \lzg(\valp_i) \,,
\end{equation}
where $\valp$ is the set of reconstructed variables used in the likelihood,
$\lsig(\valp|\mw,\gw)$, $\lzz(\valp)$ and $\lzg(\valp)$ are the 
likelihood distributions of the variables $\valp$ in WW, \zz\
and \zg\ events, and $\psig$, $\pzz$ and $\pzg$ are the (fixed) fractions
of WW, ZZ and \zg\ events in the sample, estimated from Monte Carlo 
simulation at
the corresponding centre-of-mass energy. The signal probability distribution
$\lsig(\valp|\mw,\gw)$ is obtained by reweighting four-fermion Monte
Carlo events with a true W mass of $\mwz=80.33$\,GeV and a width of 
$\gwz=2.09$\,GeV by the
ratio of two Breit-Wigner functions. The weight $f_i$ for a Monte Carlo
event $i$ with true event-by-event W boson masses \mta\ and \mtb\ is given by
\begin{equation}
f_i=\frac{B(\mta|\mw,\gw)\ B(\mtb|\mw,\gw)}
{B(\mta|\mwz,\gwz)\ B(\mtb|\mwz,\gwz)}
\end{equation}
where the Breit-Wigner function $B(m|\mw,\gw)$ 
is given by Equation~\ref{e:bwig}.
The effect of background is accounted for via
the background terms in Equation~\ref{e:rwlike}, whose probability 
distributions are calculated as a function of \valp\ using large
samples of unweighted background Monte Carlo events.

The probability distributions $\lsig$, $\lzz$ and $\lzg$ are calculated
in bins of the reweighting fit variables $\valp$, the bin size varying with
$\valp$ in order to achieve an approximately constant number of events per
bin and minimise fluctuations from limited Monte Carlo statistics. 
The likelihood for the whole sample is therefore obtained from the
number of events $N_j$ in each bin $j$ where the variables take the values
$\valp_j$:
\begin{equation}
\ltot(\mw,\gw)=\prod_j \left[ \left( \levt(\valp_j|\mw,\gw)\right)^{N_j} 
\right]
\end{equation}

Two types of reweighting fit are performed. In the first, 
the likelihood $\ltot$ is maximised as \mw\ is varied
and \gw\ is determined from \mw\  by the Standard Model relation 
(Equation~\ref{e:smwidth}).
In the second, a two parameter fit is performed, allowing both \mw\ and \gw\
to vary simultaneously. The results for \mw\ are very similar in both 
cases, but for consistency with the convolution and Breit-Wigner analyses,
the result of the first fit is used for the reweighting fit
W-mass result in this paper.

\subsection{The \bqqlv\ reweighting fit}

The \qqlv\ reweighting fit uses the same basic event selection as the
convolution fit. In the \qqev\ and \qqmv\ channels, both 4C and 5C 
kinematic fits are performed, and all events
for which both kinematic fits converge are retained.
The reweighting fit is performed simultaneously in three reconstructed 
variables which make up the variable set $\valp$:
\begin{itemize}
\item The reconstructed W mass from the 5C fit, $\mvc$, in 16 bins from~65 
to~105\,GeV.
\item The error on the reconstructed 5C fit mass, $\sigma_{\mvc}$, in five bins
from 0.5 to~6.5\,GeV.
\item The two-jet invariant mass from the 4C fit, in four bins
from 40 to 140\,GeV.
\end{itemize}
The bin sizes vary, and are chosen such that each of the 320 bins 
in each channel is populated by about 400 Monte Carlo events. Events are
discarded if any of the variables fall outside the bin ranges. The use
of the error on the 5C~fit mass and the jet-jet invariant mass significantly
improves the statistical precision of the fit as compared to the
one-dimensional reweighting using $\mvc$ alone \cite{wmass189}.

In the \qqtv\ channel, all information comes from the hadronic system and
is extracted using an analytic implementation of the 5C fit. Two variables
are used, namely the 5C fit mass (20~bins from 65 to 105\,GeV) and its
error (five bins from 0 to 6.5\,GeV). A variable bin size is again used,
with around 1000~Monte Carlo events per bin. Events with a kinematic fit 
probability of less than $10^{-3}$ are removed.

The reweighting fit technique should implicitly correct for all effects which 
bias the reconstructed W mass, providing they are
included in the Monte Carlo simulation used to generate the template 
distributions. Therefore, the effects of initial-state radiation, 
event selection and reconstruction biases are all included.
This is checked using large Monte Carlo samples over the full range of 
centre-of-mass energies and true W masses from 79.33--81.33\,GeV.
The errors returned by the fits are similarly checked by 
studying the pull distributions in Monte Carlo subsamples, and found to
be unbiased.

\subsection{The \bqqqq\ reweighting fit}

The \qqqq\ reweighting fit uses the same basic event selection as the 
convolution fit. However, the method used for the assignment of jets
to the two W bosons is rather different, with only one combination
per event entering the final fit. The tracks and clusters of the event
are first grouped into five jets using the Durham jet algorithm, 
and a 4C kinematic fit is performed. The 
value of the variable $\bar{y}_{ij}=E_iE_j(1-\cos\theta_{ij})$
calculated for each pair of jets, where $E_i$ and $E_j$ are the fitted
energies of jets $i$ and $j$ and $\theta_{ij}$ is the angle between them.
The five fitted jets are assigned to the two W bosons requiring that 
the pair of jets with the lowest $\bar{y}_{ij}$ is always kept together, both
jets being assigned to the same W boson. Each of the other three jets is
then assigned in turn to the same W boson as the paired jets.
This results in three 
distinct jet assignment combinations, which are each fitted with a 5C kinematic
fit.

In the \qqqq\ analysis with the \syaiii\ jet direction reconstruction
method, the best of the three
jet combinations is determined using the jet-pairing likelihood 
technique described in \cite{wmass189}, with the two jets corresponding to
the minimum $\bar{y}_{ij}$ merged into a single jet. For each of the possible
jet pairing assignments, three input variables are calculated and 
fed into a likelihood discriminant, and the combination with the
largest output value is retained for the fit. The likelihood reference 
distributions are determined using large Monte Carlo samples, separately
at each centre-of-mass energy. The input variables
consist of the value of the CC03 matrix element for W-pair production
\cite{wwxsec189},
determined from the measured four-vectors of the reconstructed jets;
the difference in reconstructed masses of the two W bosons, determined
using the initial 4C kinematic fit; and the sum of the di-jet opening angles.
The CC03 matrix element is averaged over three assumed W mass values
from 80.1 to 80.6\,GeV.
This algorithm selects a jet assignment combination in every selected
\qqqq\ event, and is correct 72\,\% of the time.\footnote{In this case,
each jet is associated to the original quark closest to it in angle,
and the jet assignment is considered correct if all quarks associated
to jets assigned to one W boson do in fact originate from the decay of 
one boson.}
For the \qqqq\ analysis
with the $J_0$ jet algorithm,  the jet angular resolution is
such that the CC03 matrix element provides good discriminating 
power by itself, and the jet-pairing likelihood is not used. This algorithm
selects the correct jet assignment in 74\,\% of cases.

Having selected one jet assignment combination, the corresponding 4C and 5C 
kinematic fits 
are used to provide the reconstructed variables entering the 
reweighting fit likelihood. These variables are:
\begin{itemize}
\item The reconstructed W mass from the 5C fit, $\mvc$, in 24~bins
from 65 to 105\,GeV.
\item The error on the 5C fit mass, $\sigma_{\mvc}$, in 5~bins from
zero to 5\,GeV.
\item The difference of the two 4C fit masses, $\delta m$, in 5~bins
from $-50$ to 55\,GeV.  The mass difference is signed such that the
W boson containing the jet with the highest energy before
the kinematic fit contributes with a positive sign.
\end{itemize}
As for the \qqlv\ fit, the bin sizes are chosen so that each bin is populated
by around 400~Monte Carlo events.  Events in which either of the fits fail,
or in which any of the reconstructed
variables fall outside the given range, are discarded.

The fit method is checked using large Monte Carlo samples as for the
\qqlv\ reweighting fit.
The errors returned by the fit are also checked by studying pull distributions
in Monte Carlo subsample tests, and found to be unbiased.

\subsection{Reweighting fit results}

The reweighting fit is used to analyse the data from each year and channel
separately, and the results for the different years are combined to give
the values shown in Tables~\ref{t:mresstat} and~\ref{t:wresstat}. The
results from each year are also shown separately as the `RW' points 
in Figures~\ref{f:mresyear} and~\ref{f:wresyear}. As discussed above
the mass values are determined using a one parameter fit to \mw\ only,
and the width values are determined using a two parameter fit to \mw\ 
and \gw. In the latter fits, the correlation coefficients between \mw\
and \gw\ are $0.08$ in the \qqlv\ channel and $0.07$ in the \qqqq\ channel,
and the mass results agree with those from the one parameter fits to within
1\,MeV.
No separate \gw\ results are shown for the individual \qqev, \qqmv\ and \qqtv\ 
channels in 1996 due to the small numbers of selected events, but the 1996
data are included in the overall averages.
The expected statistical errors are also given in Tables~\ref{t:mresstat}
and~\ref{t:wresstat}, evaluated using Monte Carlo subsample tests. The
statistical errors on the data results are again consistent with those expected
from Monte Carlo, taking into account the expected level of statistical 
fluctuations. 

The reconstructed mass distributions from the 5C fits can be seen in
Figures~\ref{f:mwrwbw}(a) and (b), for both the \qqlv\ and \qqqq\ 
channels (for the latter only the selected
jet assignment combinations are shown). The reweighted
Monte Carlo template distributions corresponding to the fitted values
of \mw\ in each channel are also shown, including both signal (WW events
with the correct jet assignment) and background contributions. The
width of the \qqlv\ mass peak is smaller than that from the convolution
fit shown in Figure~\ref{f:qqlvmasscv} because the latter
displays the average of the two fitted W masses in each event. This average
does not
take into account the better resolution of the \qqbar\ system mass compared
to that of the $\ell\nu$ system, information which is however included in the
convolution fit itself.

\epostfig{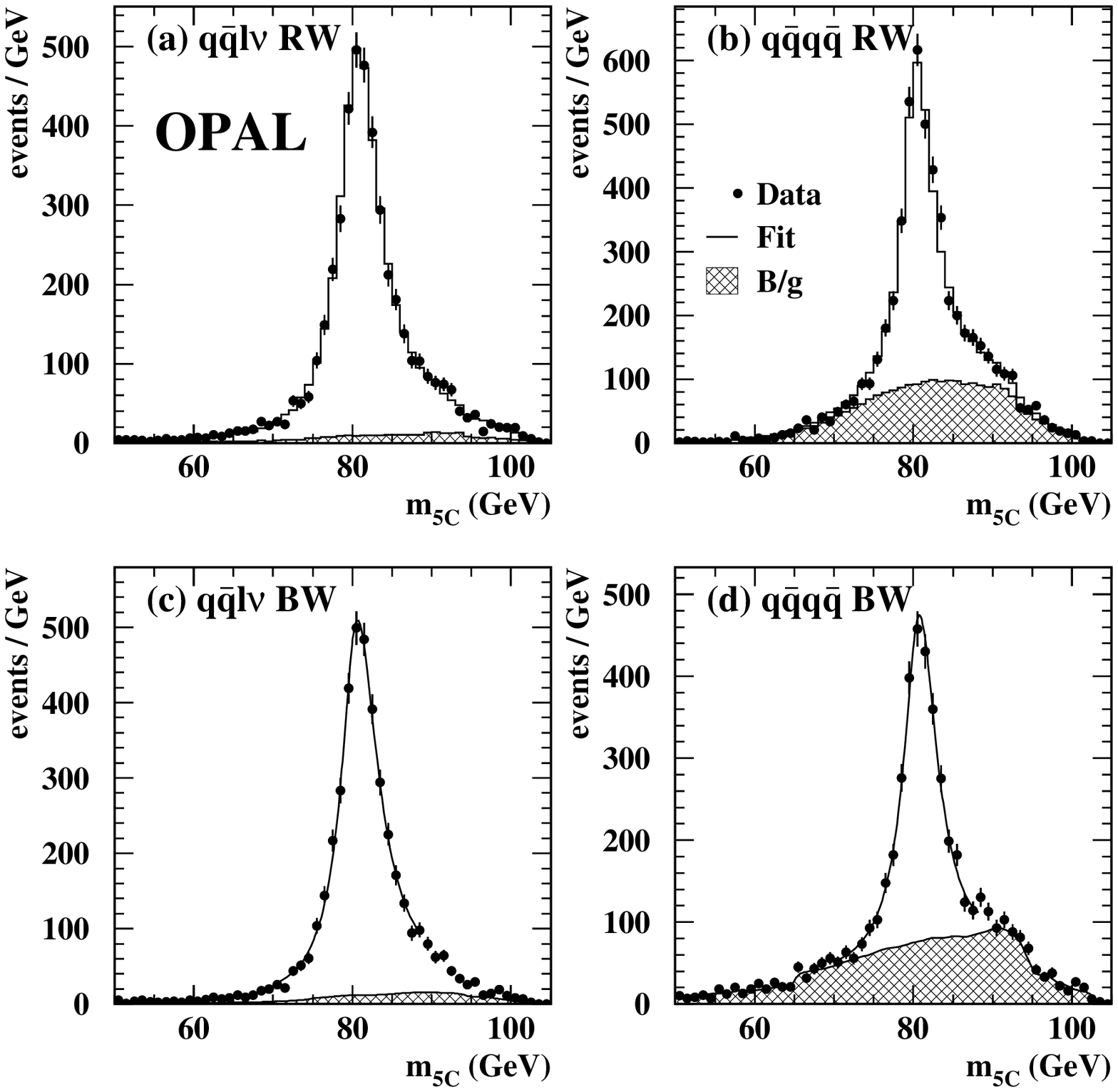}{f:mwrwbw}{Distributions of 5C fit masses
for the \qqlv\ and \qqqq\ channels in the reweighting (a,b) and Breit-Wigner 
(c,d) fits. For the reweighting fits, the histograms show the reweighted
template distributions corresponding to the fitted W mass values, and for
the Breit-Wigner fits, the fitted functions are indicated by the solid lines
drawn over the fit regions (70--88\,GeV). In both cases, the estimated
background contributions are indicated by the shaded regions.}

\section{The Breit-Wigner fit}\label{s:bwfit}

The Breit-Wigner fit is based on a simple
likelihood fit to the distribution of W boson masses reconstructed
using a 5C kinematic fit in each event, and is very similar to that described
in \cite{wmass172}. The main motivation for this analysis is to extract
the W mass using a simple and transparent method, to act as a cross-check
for the convolution and reweighting fits. The Breit-Wigner fit does not
measure the W width.

The event selection and reconstruction are very similar
to those of the convolution and reweighting fits. In the \qqlv\ channel,
only events with a 5C kinematic fit probability exceeding $10^{-3}$ 
are used in the analysis. Events in each of the lepton sub-channels
(\qqev, \qqmv\ and \qqtv) are treated separately, and 
the \qqtv\ channel is further divided into events where the $\tau$ decays
leptonically or hadronically. In the \qqqq\ channel, events are reconstructed
as five jets if the Durham jet resolution parameter $y_{45}>0.0037$
(about 23\,\% of the events), and as four jets otherwise. 
In four-jet events, 5C kinematic fits are performed
on all three possible jet pairings. The fit with the highest probability
$P_1$ is used if $P_1>0.003$ for the \syaiii\ jet direction reconstruction
method,  and $P_1>0.01$ for the $J_0$ method.
The fit with the second-highest probability $P_2$
is also used (with equal weight) if it passes both the previous probability
cut  and $P_2>\frac{1}{3}P_1$; this
occurs in approximately 20\,\% of events. In five-jet events, at most one
of the possible ten jet assignment combinations is used, selected
according to the output of the jet assignment likelihood algorithm used in
\cite{wmass189}. The 
likelihood inputs are the difference between the two W masses in a 4C fit,
the largest inter-jet opening angle 
between jets in the three-jet system, and the cosine of the polar angle of
the three-jet system. The jet combination giving the largest likelihood
value is used provided the value is greater than a minimum cut 
requirement, which happens in 73\,\% of selected five-jet WW events.

In all channels, the fitted W mass value is extracted using an unbinned
maximum likelihood fit to the distribution of reconstructed 5C fit masses
$m$ in the region $70<m<88$\,GeV. The fit function is chosen empirically
and consists of two terms:
$S(m)$ describes the signal contribution and $B(m)$ the 
combinatorial and non-WW background.
In the \qqlv\ channel, the signal function consists of an
asymmetric relativistic Breit-Wigner function with different widths
above and below the peak:
\begin{equation}
S(m)=A\frac{m^2\Gamma^2_{1,2}}{(m^2-m_0^2)^2+m^2\Gamma^2_{1,2}}\,,
\end{equation}
where $m_0$ is the fitted mass and $\Gamma_{1,2}$ are fixed parameters, 
$\Gamma_1$ being taken for $m<m_0$ and $\Gamma_2$ otherwise, and
$A$ is a normalisation constant.
In the \qqqq\ channel the signal function $S(m)$ is additionally multiplied
by a Gaussian function of mean $m_0$ and width $\sigma$, since this is found
to improve the description of the reconstructed 5C fit mass distribution.
These parameterisations were found to give adequate descriptions of 
the reconstructed distributions in Monte Carlo simulated data samples
of around ten times the data luminosity.
The parameters $\Gamma_1$ and $\Gamma_2$ were
determined using large samples of Monte Carlo signal events with 
$\mw=80.33$\,GeV, and were parameterised as linear functions of \rs.
The parameter $\sigma$ for the \qqqq\ channel was similarly determined and 
found to be independent of \rs.

The contributions from combinatorial WW background, ZZ and \zg\ final 
states are represented by the background function $B(m)$, derived from
Monte Carlo simulated events separately at each centre-of-mass energy.
The fractions of background assumed in the fits
are fixed to those observed in Monte Carlo.
As for the convolution fit, the fitted mass $m_0$ must be corrected
for biases arising from initial-state radiation, the event selection,
reconstruction and fitting procedures. Studies using Monte Carlo samples
with the full range of \rs\ values and true W masses from 79.33--81.33\,GeV
show these biases to have magnitudes of up to 500\,MeV (in the
\qqtv\ and \qqqq\ channels), to be independent of \rs, but to 
depend slightly on the true
W mass. They were therefore parameterised as linear functions of the fitted
mass using large Monte Carlo samples of both signal and background events,
and applied as corrections to the raw fitted mass values.

The Breit-Wigner fit is applied separately to the data taken at each 
centre-of-mass energy from 183\,GeV to 209\,GeV 
(dividing the 1999 and 2000 data samples into four 
and two energy bins respectively), and the results combined. The results
for each channel (including both the modified \syaiii\ and  $J_0$ 
jet direction reconstruction methods in the \qqqq\ channel) 
are shown in Table~\ref{t:mresstat}, together with
the expected statistical errors evaluated using Monte Carlo subsample tests.
The results are also shown as a function of data-taking year as the `BW'
points in  
Figure~\ref{f:mresyear}. Data taken in 1996 at $\rs\approx 172\,$GeV
have not been reanalysed, and the Breit-Wigner fit results from
\cite{wmass172} are shown. The reconstructed 5C mass distributions for 
\qqlv\ events and selected jet assignment combinations in \qqqq\ events
are shown in Figure~\ref{f:mwrwbw}(c) and~(d), together with the 
fitted functions used to extract the W mass.

\section{Systematic uncertainties}\label{s:syst}

The main systematic uncertainties in the measurements of the W mass and width
arise from the understanding of the detector calibration and performance,
the hadronisation of quarks into jets, possible final-state interactions
in the \qqqq\ channel, the modelling of non-WW background, 
the simulation of photon radiation in WW events and uncertainties in 
the LEP beam energy. These and other small systematic effects have been
calculated separately for the \qqlv\ and \qqqq\ channels, using all three
analysis techniques for the W mass, and for the convolution and reweighting
fits for the W width. The determination of 
all systematic errors is described in detail below, and the results are 
summarised in Tables~\ref{t:mwsyst} and~\ref{t:gwsyst}. Detector-related
effects tend to increase slightly with energy; other uncertainties
are taken to be constant unless stated otherwise.
The magnitudes of the systematic uncertainties are generally rather similar
between the different fitting techniques, but there are some significant
differences, and these are also discussed below.

\begin{table}
\vspace{-4mm}

\hspace{-5mm}
\begin{tabular}{l|rrr|rrr|rr|r||r}
 & \multicolumn{3}{c|}{\qqlv} & \multicolumn{3}{c|}{\qqqq} &
\multicolumn{2}{c|}{\qqqq} & Comb. & \qqqq \\
 & & & & \multicolumn{3}{c|}{\syaiii} & $J_0$ & \syavi\ & &  $\Delta m$ \\
Source & CV & RW & BW & CV & RW & BW & CV & CV & CV & CV \\ \hline
Jet energy scale & 7 & 1 & 2 & 4 & 4 & 4 & 5 & 4 & 6 & {\em 0} \\
Jet energy resolution & 1 & 1 & 1 & 0 & 1 & 3 & 1 & 0 & 0 & {\em 0}\\
Jet energy linearity & 9 & 9 & 12 & 2 & 2 & 4 & 2 & 1 & 6 & {\em 1} \\ 
Jet angular resolution & 0 & 0 & 0 & 0 & 0 & 0 & 0 & 0 & 0 & {\em 1} \\
Jet angular bias & 4 & 4 & 4 & 7 & 7 & 6 & 6 & 7 & 5 & {\em 1} \\
Jet mass scale & 10 & 7 & 6 & 5 & 11 & 3 & 5 & 5 & 8 & {\em 0} \\
Electron energy scale & 9 & 6 & 8 & - & - & - & - & - & 6 &  - \\
Electron energy resolution & 2 & 2 & 6 & - & - & - & - & - & 1 & - \\
Electron energy linearity & 1 & 1 & 2 & - & - & - & - & - & 1 & - \\
Electron angular resolution & 0 & 0 & 0 & - & - & - & - & - & 0 & - \\
Muon energy scale & 8 & 7 & 7 & - & - & - & - & - & 6 & - \\
Muon energy resolution  & 2 & 2 & 3 & - & - & - & - & - & 1 & - \\
Muon energy linearity & 2 & 2 & 2 & - & - & - & - & - & 1 & - \\
Muon angular resolution & 0 & 0 & 0 & - & - & - & - & - & 0 & - \\
WW event hadronisation & 14 & 8 & 16 & 20 & 26 & 18 & 6 & 19 & 16 & {\em 40} \\
Colour reconnection & - & - & - & 41 & 41 & 32 & 125 & 228 & 14  & - \\
Bose-Einstein correlations & - & - & - & 19 & 18 & 21 & 35 & 64 & 6 &{\em 45}\\
Photon radiation & 11 & 11 & 10 & 9 & 8 & 8 & 9 & 9 & 10 & {\em 0} \\
Background hadronisation & 2 & 1 & 2 & 20 & 12 & 32 & 17 & 24 & 8 & {\em 4} \\
Background rates & 1 & 0 & 5 & 6 & 2 & 7 & 4 & 7 & 3 & {\em 0} \\
LEP beam energy & 8 & 9 & 9 & 10 & 11 & 10 & 10 & 10 & 9 & - \\
Modelling discrepancies & 4 & 0 & 0 & 15 & 0 & 0 & 10 & 11 & 8 & {\em 5} \\
Monte Carlo statistics & 2 & 3 & 3 & 2 & 3 & 3 & 2 & 2 & 2 & {\em 3} \\
\hline
Total systematic error & 28 & 22 & 29 & 58 & 56 & 56 & 133 & 240 & 32 &{\em 61} \\
Statistical error & 56 & 58 & 64 & 60 & 64 & 73 & 51 & 73 & 42 & {\em 68}  \\
Total error & 63 & 62 & 70 & 83 & 85 & 92 & 142 & 251 & 53 & {\em 91} \\
\end{tabular}
\caption{\label{t:mwsyst}Summary of systematic uncertainties (in MeV) on the 
measurements of the W mass. Results are given separately for the
\qqlv\ and \qqqq\ channels (\syaiii\ jet direction reconstruction) with the 
convolution, reweighting and Breit-Wigner fitting methods. Results
are also given for the convolution fit for the $J_0$ and
\syavi\ jet direction reconstruction methods, and for the combination
of convolution \qqlv\ and \qqqq\ (\syaiii) results (where the combination 
takes the systematic uncertainties and their correlations into account). The 
last column gives the systematic uncertainties for the mass difference 
\dmyx{\syaiii} discussed in Section~\ref{s:fsilim}.}
\vspace{-3mm}

\end{table}

\begin{table}
\centering

\begin{tabular}{l|rr|rr|r}
 & \multicolumn{2}{c|}{\qqlv} & \multicolumn{2}{c|}{\qqqq} & Comb. \\
Source & CV & RW & CV & RW & CV \\ \hline
Jet energy scale & 0 & 7 & 0 & 2 & 0 \\
Jet energy resolution & 16 & 12 & 4 & 5 & 12 \\
Jet energy linearity & 6 & 1 & 1 & 1 & 4 \\
Jet angular resolution & 2 & 3 & 4 & 3 & 2 \\
Jet angular bias & 2 & 2 & 0 & 3 & 1 \\
Jet mass scale & 6 & 2 & 1 & 7 & 4 \\
Electron energy scale & 7 & 2 & - & - & 4 \\
Electron energy resolution & 27 & 40 & - & - & 18 \\
Electron energy linearity & 0 & 0 & - & - & 0 \\
Electron angular resolution & 1 & 0 & - & - & 1 \\
Muon energy scale & 7 & 5 & - & - & 4 \\
Muon energy resolution & 8 & 20 & - & - & 5 \\
Muon energy linearity & 1 & 1 & - & - & 0 \\
Muon angular resolution & 0 & 0 & - & - & 0 \\
WW event hadronisation & 77 & 55 & 68 & 98 & 74 \\
Colour reconnection & - & - & 151 & 136 & 53 \\
Bose-Einstein correlations & - & - & 32 & 13 & 11 \\
Photon radiation & 11 & 34 & 10 & 26 & 11 \\
Background hadronisation & 10 & 10 & 32 & 46 & 18 \\
Background rates & 18 & 20 & 34 & 32 & 24 \\
LEP beam energy & 3 & 1 & 2 & 1 & 3 \\
Modelling discrepancies & 4 & 0 & 25 & 0 & 11 \\
Mass-width coupling & 24 & 0 & 0 & 0 & 16 \\
Monte Carlo statistics & 9 & 12 & 8 & 12 & 9 \\
\hline
Total systematic error & 91 & 85 & 177 & 180 & 102 \\
Statistical error & 135 & 131 & 112 & 130 & 96 \\
Total error & 163 & 156 & 209 & 222 & 140 \\
\end{tabular}
\caption{\label{t:gwsyst}Summary of systematic uncertainties (in MeV) on the 
measurements of the W width. Results are given separately for the
\qqlv\ and \qqqq\ channels with the 
convolution and reweighting 
fitting methods.  The systematic uncertainties for the combination of
\qqlv\ and \qqqq\ convolution fit results are also shown.}
\end{table}

\subsection{Detector calibration and simulation}\label{s:detsys}

The Monte Carlo descriptions of the jet and lepton energy scales, and
energy and angular resolutions, are checked using samples of 
$\zb\rightarrow\qqbar$, $\zb\rightarrow\epem$ and $\zb\rightarrow\mpmm$
events recorded at $\rs\approx 91$\,GeV
at the beginning of each data-taking year, and at other
times during the data-taking periods in 1998, 1999 and 2000. Detailed 
comparisons between these data and corresponding Monte Carlo simulations are 
used to derive small adjustments at the level of reconstructed 
jets and leptons, which are then applied to the WW and background
samples that are used to calibrate the bias corrections in the convolution and 
Breit-Wigner fits and to derive template distributions in the reweighting
fit. Where necessary, the energy and angular resolution in the Monte Carlo
simulation are degraded by applying Gaussian smearing to the reconstructed
quantities. The associated systematic uncertainties are propagated to the 
measured W mass and width by varying the applied corrections.

The properties of jets are checked using samples of two-jet and three-jet 
$\zb\rightarrow\qqbar$ events, and $\zg\rightarrow\qqbar$ events
from the high-energy LEP2 data samples, as follows:
\begin{description}
\item[Jet energy scale:] This is checked using $\zb\rightarrow\qqbar$ 
events reconstructed as
two jets with the Durham algorithm and satisfying $y_{23}<0.02$. The
same particle selection requirements and energy double-counting 
correction procedure
are applied as for the WW analysis. The mean of the sum of the two
jet energies is studied as a function of 
$\cos\theta=\frac{1}{2}(|\cos\theta_1|+|\cos\theta_2|)$, where $\theta_1$
and $\theta_2$ are the reconstructed polar angles of the two jets.
The ratio of these energy sums in data and Monte Carlo is shown in 
Figure~\ref{f:wwfixj}(a), and is used to derive corrections to the Monte Carlo
energy scale as functions of jet $\cos\theta$ and data-taking year. 
The corrections in the forward region
beyond $\cos\theta=0.85$ are much larger than in the central region, due to
the difficulties in accurately modelling the complex detector geometry and 
larger amount of dead material. The residual
uncertainty on the jet energy scale is 0.4\,\%, dominated by contributions
from \zb\ data statistics, possible quark-flavour dependences (assessed by
repeating the studies after removing events with a reconstructed secondary
vertex indicating a heavy quark decay \cite{oldrb}) and possible variations 
during the course of a year. 

\epostfig{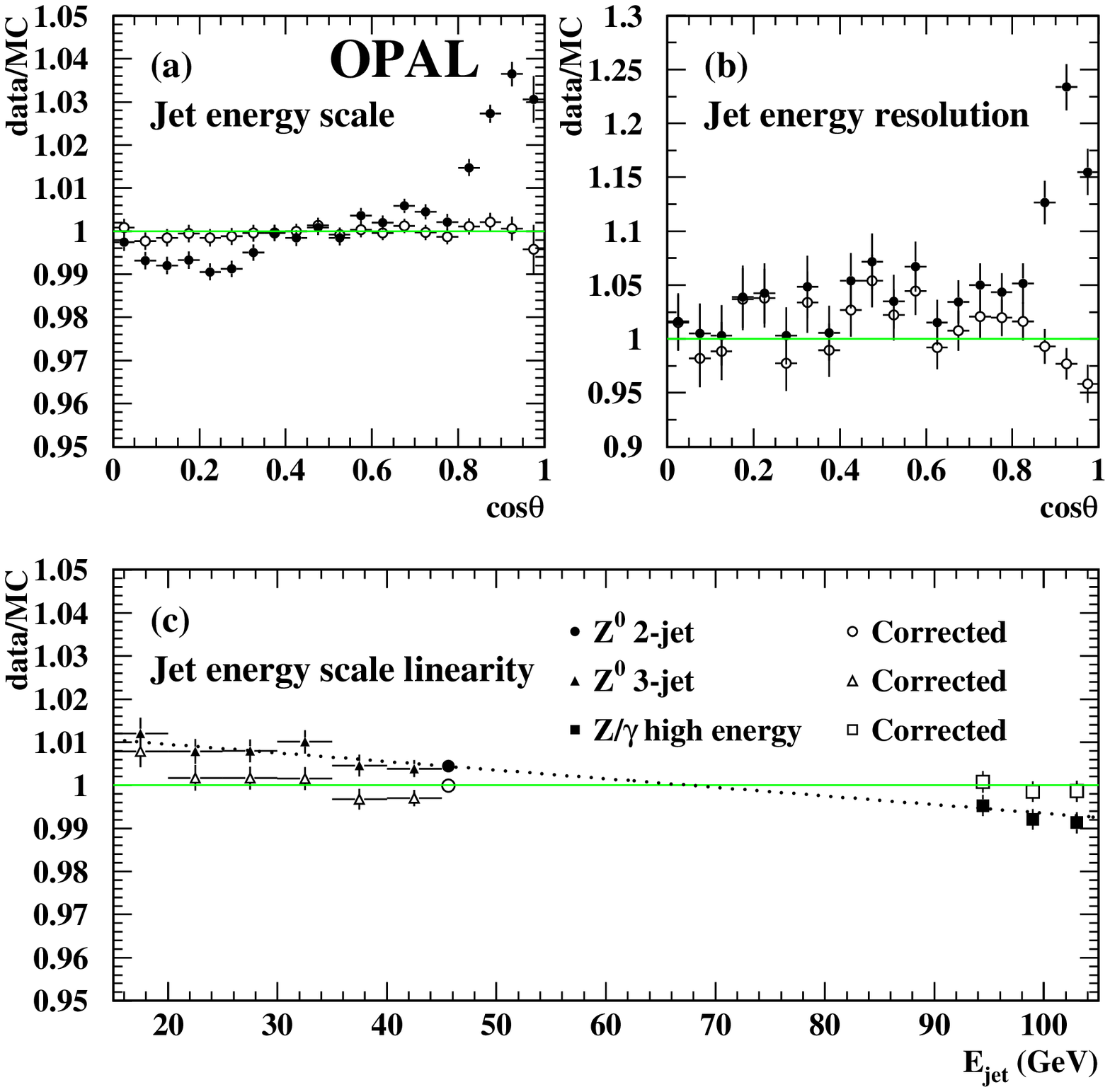}{f:wwfixj}{Determination of energy corrections 
for jets (see text). Ratios of data to Monte Carlo are shown averaged over
all data-taking years for: 
(a) jet energy scale as a function
of $\cos\theta$, (b) jet energy resolution as a function of $\cos\theta$,
(c) jet energy scale as a function of the jet energy itself. The results
using the uncorrected simulation are shown by the filled points, and those
with the corrected simulation are shown by the open points, with the
error bars indicating the statistical error in each case. The horizontal
lines indicate ratios of unity, and the dotted line in (c) shows the
linearity correction used to parameterise the jet energy scale dependence
on the jet energy itself.}

\item[Jet energy resolution:] The width of the distribution of two-jet
energy sums is sensitive to the jet energy resolution, and was studied
using the same techniques as the energy scale. The ratio of widths seen in
data and Monte Carlo is shown in Figure~\ref{f:wwfixj}(b)---the Monte Carlo
resolution is about 4\,\% too good for $\cos\theta<0.85$, and up to 20\,\%
too good in the forward region beyond $\cos\theta=0.85$. After correction,
the residual uncertainty lies between 0.6\,\% and 2\,\% depending on 
$\cos\theta$, limited by \zb\ data statistics.

\item[Jet energy linearity:] The studies with two-jet $\zb\rightarrow\qqbar$
events check the energy scale for $\sim 45$\,GeV jets, close to the 
average energy of jets produced in W decays, but event-by-event the latter
range from about 20\,GeV to 85\,GeV. It is therefore
important to check the linearity of the energy response, {\em i.e.\/} the
energy scale for lower and higher energy jets. This has been studied by 
looking both at $\zb\rightarrow\qqbar\rm g$ three-jet events and high
energy $\zg\rightarrow\qqbar$ two-jet events. 
Coplanar three-jet events are selected by
requiring $y_{23}>0.02$ and $y_{34}<0.005$, and that the sum of
the inter-jet angles exceeds 355$^\circ$. The jet energies can then be computed
using the measured jet angles and masses, and the ratio of reconstructed
to expected energies determined as a function of expected energy. The ratio
of this quantity in data and Monte Carlo is shown in Figure~\ref{f:wwfixj}(c),
from which it can be seen that the energy scale in data is around
0.5\,\% higher for 30\,GeV jets than for 45\,GeV jets. 

The behaviour at high jet energies is studied with $\zg\rightarrow\qqbar$
events taken at $\rs>180\,$GeV, and satisfying $y_{23}<0.02$
and $\sqrt{s'/s}>0.85$ where the reconstructed \epem\ collision energy
after any initial-state radiation \rsp\ is calculated as in 
\cite{opalff}.
In these events, the behaviour of the jet energy scale as a function of
$\cos\theta$ is consistent with that seen for 45\,GeV jets, but the
overall energy scale is shifted downwards by about 1\,\%, as can be seen
for the high jet energy points in Figure~\ref{f:wwfixj}(c).

These studies are consistent with a linear dependence of the jet energy 
scale on the jet energy itself, with a slope of 
$(-2.00\pm 0.30)\times 10^{-4}$. The corresponding
correction is applied to the energy scale in Monte Carlo
simulation. The uncertainty is dominated by data statistics 
($0.26\times 10^{-4}$), but also includes
systematic contributions from two-photon ($0.06\times 10^{-4}$) and 
$\tau$-pair ($0.09\times 10^{-4}$)
background modelling in the high energy \qqbar\ samples, and
possible quark flavour dependences ($0.09\times 10^{-4}$). 
Effects from hadronisation 
and four-fermion background modelling are found to be negligible.
This uncertainty on the correction contributes a systematic error of
around 4\,MeV in the \qqlv\ and 2\,MeV in the \qqqq\ W mass measurements.

Although the data are consistent
with a linear slope, a second order polynomial is also fitted and used
to correct the simulation as an alternative. The corresponding W mass
uncertainties when the curvature is varied within the range allowed by the
data are 8\,MeV and 2\,MeV in the \qqlv\ and \qqqq\ channels.
The final jet energy linearity uncertainties on the W mass
and width are calculated as the quadrature sum of the shifts resulting from
changing the linear correction by its uncertainty, and using the alternative
second order polynomial correction with the maximum curvature allowed by the
data.

\item[Jet angular resolution:] The jet $\cos\theta$ and $\phi$ resolutions
are checked by using the two-jet $\zb\rightarrow\qqbar$ sample and studying
the widths of the distributions of $\cos\theta_1+\cos\theta_2$ and 
$\phi_1-\phi_2$. These are found to be 4\,\% and 1\,\% narrower in Monte Carlo
than data for the $J_0$ jet direction reconstruction method, independent of
$\cos\theta$, and are smeared accordingly. The corresponding uncertainties
are 0.4\,\% for $\cos\theta$ and 0.3\,\% for $\phi$, dominated by 
\zb\ data statistics. The uncertainties for the \syaiii\ direction
reconstruction method are similar, though modelling of the jet angular 
resolution is somewhat worse. The differences between data and Monte Carlo 
are around a factor two larger than for the $J_0$ direction reconstruction,
necessitating correspondingly larger Monte Carlo corrections.

\item[Jet angular bias:] A bias in the jet $\cos\theta$ reconstruction, equal
in magnitude but opposite in sign for positive and negative $\cos\theta$,
would not show up in the jet acollinearity measured by 
$\cos\theta_1+\cos\theta_2$, and could have significant effects on the
W mass and width. This is studied by using individual jets in two-jet
$\zb\rightarrow\qqbar$ events, and calculating their $\cos\theta$ separately
using tracking and calorimeter information (the two detector systems have
independent and uncorrelated angular reconstruction uncertainties). Some
differences are seen, but these are generally well modelled by the Monte Carlo
simulation. Since the jet $\cos\theta$ information is determined from both
the tracking and calorimeter information, half the residual difference
between data and Monte Carlo tracking-calorimeter 
deviations is taken as a systematic uncertainty
on the absolute $\cos\theta$ bias. This study was performed for both
the unmodified and alternative jet direction reconstruction algorithms, and no 
significant differences were seen.

\item[Jet mass scale:] No useful information on jet masses can be obtained
from studies of \zb\ data alone, and there are significant differences
between the predictions of the {\sc Jetset}, {\sc Ariadne} and {\sc Herwig}
Monte Carlo models. The jet masses predicted by the Monte Carlo are therefore
left unchanged by default, and a systematic error is assessed by scaling 
and smearing them event-by-event
by the same factor as the corresponding jet energies---this is appropriate 
for the
extreme case of jets composed entirely of massless particles. As an additional
cross-check, the energies of all particles were scaled by the same
factor as the jet energies, but before calculating the jet invariant masses.
This gave results which were negligibly different from those obtained 
by scaling the jet masses.

\end{description}

A similar approach is taken to study the modelling of leptons (electrons and 
muons separately), using
$\zb\rightarrow\epem$ and $\zb\rightarrow\mpmm$ events as follows:
\begin{description}
\item[Lepton energy scale:] The lepton energy scale is studied as a function
of $\cos\theta$ using the means of distributions of lepton energies.
For electrons, the ratios of data to Monte Carlo means
are typically within 1\,\% of unity, and are used to correct the Monte Carlo
simulation, with a residual systematic uncertainty of 0.3\,\% including
contributions from possible time dependence, the comparison of two independent
event selections and data statistics. The ratios of data to Monte Carlo 
for muons
typically agree to better than 0.3\,\%, and a systematic uncertainty of
0.3\,\% is assigned, dominated by \zb\ data statistics and the comparison
of event selections.

\item[Lepton energy resolution:] Studies of the  width of the lepton energy 
distributions show that the electron energy resolution is around 19\,\%
worse in data than Monte Carlo in the barrel region and 7\,\% worse in
the endcap, and that the muon resolution
is 6\,\% worse in data. These corrections are applied to the Monte Carlo
with corresponding uncertainties of 2\,\%, dominated by data statistics.
There are also tails in the data resolution which are not well modelled by 
the Monte Carlo. For the W mass, a systematic error is conservatively estimated
by doubling the Monte Carlo resolution correction, whilst for the width (which
is much more sensitive to such tails) a more elaborate two-component smearing 
procedure is used to model both the core and tail resolution,
 with an additional uncertainty relating to the choice of 
smearing parameters.

\item[Lepton energy linearity:] Possible dependences of the lepton energy 
scale on the lepton energy itself are studied using $\epem\gamma$ and
$\mpmm\gamma$ events taken both during the \zb\ calibration and high energy
running, by comparing the measured lepton energies with those determined
from the track and cluster angles. No significant effects are seen, within
a statistical precision on the slope of $3\times 10^{-5}$ for electrons and
$6\times 10^{-5}$ for muons, and these values are used to assess the
corresponding uncertainties on the W mass and width.

\item[Lepton angular resolution:] The lepton $\cos\theta$ and $\phi$ 
resolutions are studied using the distributions of 
$\cos\theta_1+\cos\theta_2$ and $\phi_1-\phi_2$ in the same way as discussed
above for jets. No evidence for Monte Carlo mis-modelling is seen, and the
corresponding uncertainties are obtained from the statistical precision of
the tests of 5--10\,\%.

\end{description}
The effects of all these uncertainties on the various W mass and width
analyses are shown in Tables~\ref{t:mwsyst} and~\ref{t:gwsyst}. 
Uncertainties which affect only the \qqev\ or \qqmv\ sub-channels are 
given in terms of their effect on the combined \qqlv\ results, and the
uncertainties due to the measurements of jets and leptons are assumed
to be uncorrelated. For the
W mass, the most significant uncertainties are those associated with 
energy scales, jet angular biases and jet masses, whilst for the W width,
energy resolution uncertainties play a bigger role. The uncertainties
for the three fit methods are rather similar, except for the jet energy
scale uncertainty in the \qqlv\ channel
which is significantly smaller in the reweighting  and
Breit-Wigner fits than in the convolution fit. This is due to the different
sensitivities of 4C and 5C kinematic fits with and without lepton information
to variations of the jet energy scale, and the different use made by 
the three analysis methods of the different types of fit.

A further correction is applied for the effects of beam-related background
and detector noise, which lead to additional non-simulated occupancy in
the detectors. This correction is evaluated by superimposing data events 
taken with a random beam-crossing trigger onto Monte Carlo WW and background
events. The most significant effects come from additional clusters in the 
calorimeters (especially the hadron calorimeter in the forward region), 
and lead to shifts of around 10\,MeV in the W mass and 2\,MeV in the W width, 
with systematic
uncertainties which are negligible in comparison to other detector-related
effects. The effect of this noise on the data {\em vs.\/} Monte Carlo 
comparisons discussed above was also checked and found to be negligible.

\subsection{Hadronisation in $\rm\boldmath W\rightarrow \qqbar$ decays}
\label{s:hadrn}

Uncertainties due to hadronisation in $\rm W\rightarrow\qqbar$ decays
are studied using large Monte Carlo samples where the same original 
four-fermion events have been hadronised using various different
Monte Carlo models (string, colour dipole and cluster, as implemented
in {\sc Jetset}, {\sc Ariadne} and {\sc Herwig} respectively) and 
parameter sets (see Section~\ref{s:dmc}). These models have all been 
tuned to give a reasonable overall
description of OPAL or ALEPH $\zb\rightarrow\qqbar$ data. Different models
and tuned parameter sets describe particular features of the data to a
greater or lesser extent, reflecting partly the emphasis placed
on various
variables ({\em e.g.} event shapes, charged and neutral particle 
multiplicities and fragmentation functions) by the different tune 
procedures.

All these models give adequate descriptions of general event properties
in $\ww\rightarrow\qqlv$ and $\ww\rightarrow\qqqq$ events, and the limited
data statistics do not allow any of the models to be disfavoured or excluded.
However, they predict different fit biases or reweighting template 
distributions for the W mass and width fits. This is illustrated in 
Table~\ref{t:hadrm}, which shows the biases in fitted W mass and width
from the convolution fit analysis applied to the same event samples hadronised
with various different models, but calibrated using Monte Carlo events 
hadronised using JT. The statistical errors on the mass differences
are calculated using a Monte Carlo subsample technique and take into account
the correlation between the samples due to the common initial four-fermion
events. Taking the `raw' mass shifts from Table~\ref{t:hadrm}, it is clear
that the different models and tunes 
predict significantly different W mass biases,
of up to $\sim 40$\,MeV in both \qqlv\ and \qqqq\ channels, and 
corresponding width biases of up to $\sim 80$\,MeV.

\begin{table}
\centering

\begin{tabular}{l|rrrr|rrrr}
Models & \multicolumn{4}{c|}{Raw mass shifts} & 
\multicolumn{4}{c}{Adjusted mass shifts} \\
  & \qqlv & \qqqq & \qqqq & \qqqq & \qqlv & \qqqq & \qqqq & \qqqq \\
 & & \syaiii & $J_0$ & \syavi &  & \syaiii & $J_0$ & \syavi \\ \hline
JT$'$-JT &  $-32\pm 4$ & $-32\pm 4$ & $-40\pm 3$ & $-33\pm 5$ &
 $1\pm 4$ & $4\pm 4$ & $5\pm 3$ & $15\pm 5$ \\
AR-JT  & $-25\pm 3$ & $-28\pm 4$ & $-31\pm 4$ & $-35\pm 5$ & 
 $-7\pm 4$ & $-6 \pm 4$ & $-4\pm 4$ & $-6\pm 5$ \\
AR$'$-JT & $-7\pm 4$ & $-29\pm 5$ & $-12\pm 4$ & $6\pm 5$ &
 $1\pm 4$ & $-20\pm 5$ & $2\pm 4$ & $18\pm 5$ \\
HW-JT  & $-15\pm 4$ & $6\pm 4$ & $-3\pm 3$ & $-15\pm 5$ &
 $-13\pm 4$ & $1\pm 4$ & $-3\pm 3$ & $-15\pm 5$ \\
\hline
& \multicolumn{4}{c|}{Raw width shifts} & \\
JT$'$-JT & $-35\pm 9$ & --- & $68\pm 7$ & --- & \\
AR-JT & $-43\pm 9$ & --- & $-9\pm 7$ & --- & \\
AR$'$-JT & $-77\pm 9$ & --- & $-9\pm 8$ & --- & \\
HW-JT & $-5\pm 9$ & --- & $-42\pm 8$ & --- &  \\
\end{tabular}
\caption{\label{t:hadrm} Differences in W mass and width biases (in MeV) 
evaluated using
the convolution fit for various different hadronisation models in comparison
to the default JT model. For the W mass,
 results are given for the \qqlv\ channel,
and for the \qqqq\ channel using the \syaiii, $J_0$ and \syavi\ jet direction
reconstruction methods. Results are also given after
adjusting the non-JT simulated events to have the same kaon 
and baryon content as that predicted by JT. The uncertainties
are due to finite Monte Carlo statistics, and the definitions
of the models JT, JT$'$, AR, AR$'$ and HW are given in Section~\ref{s:dmc}.}
\end{table}

The mass biases have also been studied at the `hadron' level, performing the
jet finding on all stable hadrons\footnote{Following the convention for
\zb\ decay multiplicities in \cite{pdg02}, all particles with lifetimes of
more than $3\times 10^{-10}$\,s were considered stable.} 
produced by the Monte Carlo hadronisation
model (before detector simulation), and then applying the full convolution
analysis to these jets, together with the reconstructed leptons in the 
\qqlv\ channel. The results show no significant differences between 
hadronisation models and tunes, showing that the biases are produced by 
the interplay of hadronisation and detector effects. At the hadron level,
the various models predict different average jet masses for the two
jets produced in $\rm W\rightarrow\qqbar$ decays, but these differences are 
compensated by different inter-jet angles once the decays are boosted into
the laboratory frame, where the invariant mass $m_{\rm jj}$ of the 
jet-jet system is given by:
\begin{equation}\label{e:mjj}
m_{\rm jj}^2=m_1^2+m_2^2+2E_1E_2(1-\beta_1\beta_2\cos\theta_{12})
\end{equation}
where $m_i$, $E_i$ and $\beta_i$ are the mass, energy and velocity of
jet $i$, and $\theta_{12}$ is the angle between the two jets.
The resulting average jet-jet invariant masses are 
therefore equal in all models. However, the detector-level jet reconstruction
introduces biases in both jet mass and angular distributions, and these
biases are different in the various models, spoiling the hadron level
compensation of the jet mass differences by the inter-jet angle differences
and leading to significant differences in fitted W mass between models.

A large part of the jet mass and angle biases is found to result from
deficiencies in the reconstruction of kaons and baryons. 
In jet reconstruction, all charged particles
are assigned the pion mass, and all neutral clusters zero mass. 
Charged kaons and protons (having $m>m_{\pi}$) will therefore be incorrectly
reconstructed, as will $\rm K^0_{\rm L}$ and neutrons which in addition
tend to have their
energies badly estimated in the calorimeters. Although the hadronisation models
have been tuned to \zb\ data including kaon and baryon rates, there are 
significant differences between them, with {\em e.g.}\ JT and AR 
underestimating
the production rates of kaons, and JT$'$ and AR overestimating the production
rates of baryons. Table~\ref{t:hadrm} also shows the W mass differences between
models after adjusting all the alternative models to have the same
kaon and baryon content as JT.\footnote{Since particle 
multiplicities increase with the mass of the decaying boson, the event-by-event
true W masses and decay multiplicities are correlated, leading to `artificial'
W mass shifts when kaon and baryon multiplicities are adjusted. This
effect is removed using a second iteration in the adjustment procedure, which
then changes the kaon and baryon content but leaves the average true W mass 
unchanged.}
It can be seen that, particularly for the \qqlv\ and \qqqq\ channels
with the $J_0$ jet algorithm, the differences between models are
greatly reduced after this adjustment procedure. Many other variables have
also been studied, but no other significant dependences have been found,
and the remaining mass differences are therefore taken to be indicative
of genuine differences between the hadronisation models.

In the \qqqq\ analyses with modified jet direction reconstruction, significant
mass bias differences persist after adjusting the kaon and baryon
multiplicities. This is not surprising, as these reconstruction methods
calculate the jet masses using one set of particles, and remove some
particles or weight them differently when calculating the jet angles.
This will again tend to spoil the cancellation between jet mass and angle
differences seen at hadron level (see Equation~\ref{e:mjj}), 
and the differences represent genuine
uncertainty due to the modelling of hadronisation.  The Monte Carlo models
have been tuned to reproduce inclusive \zb\ event properties, and it is not
obvious that they can also be relied on to model {\em e.g.\/} event shape
distributions when some particles are removed. To check this, event shape
distributions were calculated in $\zb\rightarrow\qqbar$ events using
only particles satisfying $p>2.5$\,GeV. The agreement between data
and the various Monte Carlo hadronisation models for these 
distributions was found to be reasonable, and not significantly worse
than for the inclusive distributions. Therefore, the models can also be 
expected to give a reasonable description of jet properties when particles
with $p<2.5$\,GeV are removed.

The final uncertainty on the W mass from the hadronisation of signal
WW events is made up of two parts: the residual differences between
hadronisation models after adjusting them to the same kaon and baryon
production rates (where the largest difference between JT and
any of the other models is taken), and an uncertainty related to the 
knowledge of kaon and baryon production rates in $\rm W\rightarrow\qqbar$
decays. The latter are calculated from the measured $\rm K^+$,
$\rm K^0_{\rm S}$ and proton production rates in \zb\ decays
of $2.242\pm 0.063$, $1.025\pm 0.013$ and $1.048\pm 0.045$ \cite{pdg02}. The 
$\rm K^0_{\rm L}$ rate is assumed to be equal to the $\rm K^0_{\rm S}$ rate
and the neutron to proton ratio to be that predicted by JT (0.97).
The ratios of kaon ($\rm K^+$ and $\rm K^0_L$) and 
baryon (proton and neutron) production in W and Z decays,
$R=n_{\rm W}/n_{\rm Z}$, are taken to be $R_{\rm K}=0.90\pm 0.01$ and 
$R_{\rm B}=0.95\pm 0.06$, where the central values are taken from 
JT and the errors from the largest
difference between JT and any other model.\footnote{
$\rm K^0_s$, $\Lambda$ and other hyperons are not included, since
they typically decay into other particles which are already accounted for.}
The measurements
and ratios were combined to give predicted production rates
of kaons and baryons in W decays of $n_{\rm K}(\rm W)=2.94\pm 0.06$
and $n_{\rm B}(\rm W)=2.00\pm 0.12$. The uncertainties on these
rates give small additional systematic errors on the W mass and width
due to the dependence of the biases on kaon and baryon production.
Small corrections are also applied to the fitted values to compensate
for the differences between these predictions and those of JT.

The differences in W width bias predicted by the various models are not
related to kaon and baryon multiplicity, and do not reduce when the
adjustment techniques used for the W mass are applied. No other variables
have been found that play a similar role to  the kaon and baryon 
multiplicities for the W mass. The largest raw difference
between JT and any of the other models is therefore used to set the systematic
uncertainty for the W width.

Taking everything into account, the final uncertainties due to 
hadronisation in signal WW events are given in Tables~\ref{t:mwsyst}
and~\ref{t:gwsyst}. The systematic errors are dominated by the residual 
differences between hadronisation models, the uncertainties on kaon and baryon
production contributing less than 5\,MeV. The full analysis has been carried 
out for all three fitting methods, which are found to have slightly 
different residual mass shifts and dependences on kaon and baryon 
multiplicities, particularly in the \qqlv\ channel where the three methods
employ 4C and 5C kinematic fits in different ways. However, the overall
hadronisation uncertainties are broadly similar.
Uncertainties due
to hadronisation in the $\zg\rightarrow\qqbar$ background are discussed
in Section~\ref{s:bgsyst} below.

\subsection{Final-state interactions}\label{s:fsi}

At LEP2 energies, the two W bosons produced in an $\epem\rightarrow\ww$
event decay when their spatial separation is about 0.1\,fm,
much smaller than the typical hadronisation scale of 1\,fm. The two
hadronising $\rm W\rightarrow\qqbar$ systems in a $\ww\rightarrow\qqqq$
event therefore overlap, and their hadronisation may involve final-state
interactions (FSI) between them, leading to exchange of four-momentum
between the decay products of the two W bosons and possible biases in the
reconstructed invariant mass spectra.
Two sources of such interactions have been
widely considered: colour reconnection and Bose-Einstein correlations
(BEC). Both of these effects are well established in other systems, but are
neither conclusively confirmed nor ruled out in $\ww\rightarrow\qqqq$ events
\cite{smew}.
The systematic uncertainties on the W mass and width from FSI effects are
therefore determined by considering various phenomenological models which
are consistent with the current limited knowledge of FSI in 
$\ww\rightarrow\qqqq$ events.

Colour reconnection effects in the perturbative phase have been shown to be
small, giving rise to possible W mass biases of around 1\,MeV \cite{crsk}.
However colour reconnection in the non-perturbative hadronisation phase
may be substantial, and the mass and width biases can only be evaluated using 
Monte Carlo models. The models of Sj\"ostrand and Khoze (SK~I, SK~II and 
SK~II$'$, as implemented in {\sc Pythia} \cite{crsk}), L\"onnblad 
(AR2 and AR3 as implemented in {\sc Ariadne} \cite{crar}), and 
{\sc Herwig} \cite{herwig} have been evaluated, and the resulting mass 
and width biases for all three analysis methods are summarised in 
Table~\ref{t:colrec}. 

\begin{table}

\hspace{-8mm}
\begin{tabular}{l|rrrrr|rr}
 & \multicolumn{5}{c|}{Mass shift (MeV)} & 
\multicolumn{2}{c}{Width shift (MeV)} \\
 & \syaiii & \syaiii & \syaiii & $J_0$ & \syavi & $J_0$ & $J_0$ \\
Model & CV & RW & BW & CV & CV & CV & RW \\ \hline
SK I ($k_I=0.9$) & $19\pm 2$ & $21\pm 2$ & $16\pm 2$ & $63\pm 2$ & $113\pm 3$
& $86\pm 5$ & $80\pm 3$  \\
SK I ($k_I=2.3$) & $41\pm 2$ & $41\pm 2$ & $32\pm 3$ & $125\pm 2$ & $228\pm 3$ &
 $150\pm 5$ & $144\pm 5$ \\
SK I ($k_I=100$) & $136\pm 3$ & $142\pm 3$ & $105\pm 4$ & $390\pm 3$ & $674\pm 4$ & 
$289\pm 6$ & $295\pm 6$ \\
SK II & $-9\pm 5$ & $24\pm 6$ & $-5\pm 3$ & $-3\pm 5$ & $0\pm 5$ & $33\pm 12$ & $26\pm 10$ \\
SK II$'$ & $-6\pm 5$ & $-1\pm 6$ & $-2\pm 4$ & $-5\pm 5$ & $4\pm 5$ & $30\pm 12$ & $29\pm 10$ \\
\hline
AR2-AR1 & $29\pm 5$ & $27\pm 6$ & $28\pm 7$ & $66\pm 5$ & $102\pm 5$ & $128\pm 11$ & $104\pm 11$ \\
AR3 & $61\pm 25$ & $41\pm 24$ & $40\pm 31$ & $145\pm 22$ & $251\pm 31$ & $348\pm 49$ & $348\pm 25$  \\
HW & $22\pm 5$ & $35\pm 8$ & $15\pm 9$ & $42\pm 5$ & $60\pm 5$ & $27\pm 11$ & 
$55\pm 15$ \\ \hline
LUBOEI BEC & $-24\pm 8$ & $-22\pm 9$ & $-27\pm 10$ & $-46\pm 7$ & $-83\pm 11$ & $41\pm 14$ 
& $17\pm 17$ \\
\end{tabular}
\caption{\label{t:colrec} W mass and width shifts (in MeV) in the \qqqq\ 
channel for various 
colour reconnection models and parameters, and for the LUBOEI Bose-Einstein
correlation model, evaluated at the mean centre-of-mass energy 
of the data sample (196\,GeV). Mass shifts are given for all three
analysis methods with the \syaiii\ jet direction reconstruction method, and for
the convolution fit with other jet reconstruction methods. The uncertainties
are due to finite Monte Carlo statistics.}
\end{table}

The mass and width shifts for the SK~I model are evaluated using Monte Carlo
samples generated with and without colour reconnection. The fraction
of events with colour reconnection depends on the SK~I strength parameter
$k_I$. For each event, the reconnected or non-reconnected
version is chosen according to a probability given by $\precon=1-\exp(-Vk_I)$,
where $V$ is the event-by-event space-time integrated product of the
maximum colour field strengths of the two overlapping strings
connecting the
two quarks in each $\rm W\rightarrow\qqbar$ decay. The value of $k_I$ is not
predicted by the model, and results for $k_I=0.9$, 2.3 and 100 are given
in Table~\ref{t:colrec}, corresponding to colour reconnection probabilities
of 0.35, 0.57 and 0.97. Both the shifts and the reconnection probability
for a given $k_I$ vary with centre-of-mass energy, {\em e.g.}
from 37\,MeV at $\rs=189$\,GeV to 47\,MeV at 207\,GeV for $k_I=2.3$ in the
convolution fit. They have therefore been
evaluated at $\rs=196$\,GeV, the mean centre-of-mass 
energy of the data sample, assuming linear dependences on \rs.
The convolution fit mass shifts for the \syaiii\, $J_0$
and \syavi\ jet direction reconstruction methods
are also shown as functions of reconnection
probability in Figure~\ref{f:colrec}, together with the 
corresponding shifts for the W width measurement.
 The measured value of $\dmyx{\syaiii} = \dmres\pm\dmstat$\,MeV (where the
error is purely statistical---see 
Section~\ref{s:cvres}) favours a reconnection probability of around 50\,\%;
further discussion of this result is given in Section~\ref{s:fsilim}.
Table~\ref{t:colrec} also gives results for the SK~II and SK~II$'$ models,
where the colour strings have infinitesimally small radii, but colour
reconnection occurs with unit probability on the first crossing (SK~II)
or only if it would reduce the total string length (SK~II$'$). Both of
these models predict smaller mass and width biases than SK~I even with 
moderate values of $k_I$, and are therefore not considered further when 
setting the systematic uncertainties on the W mass and width.

\epostfig{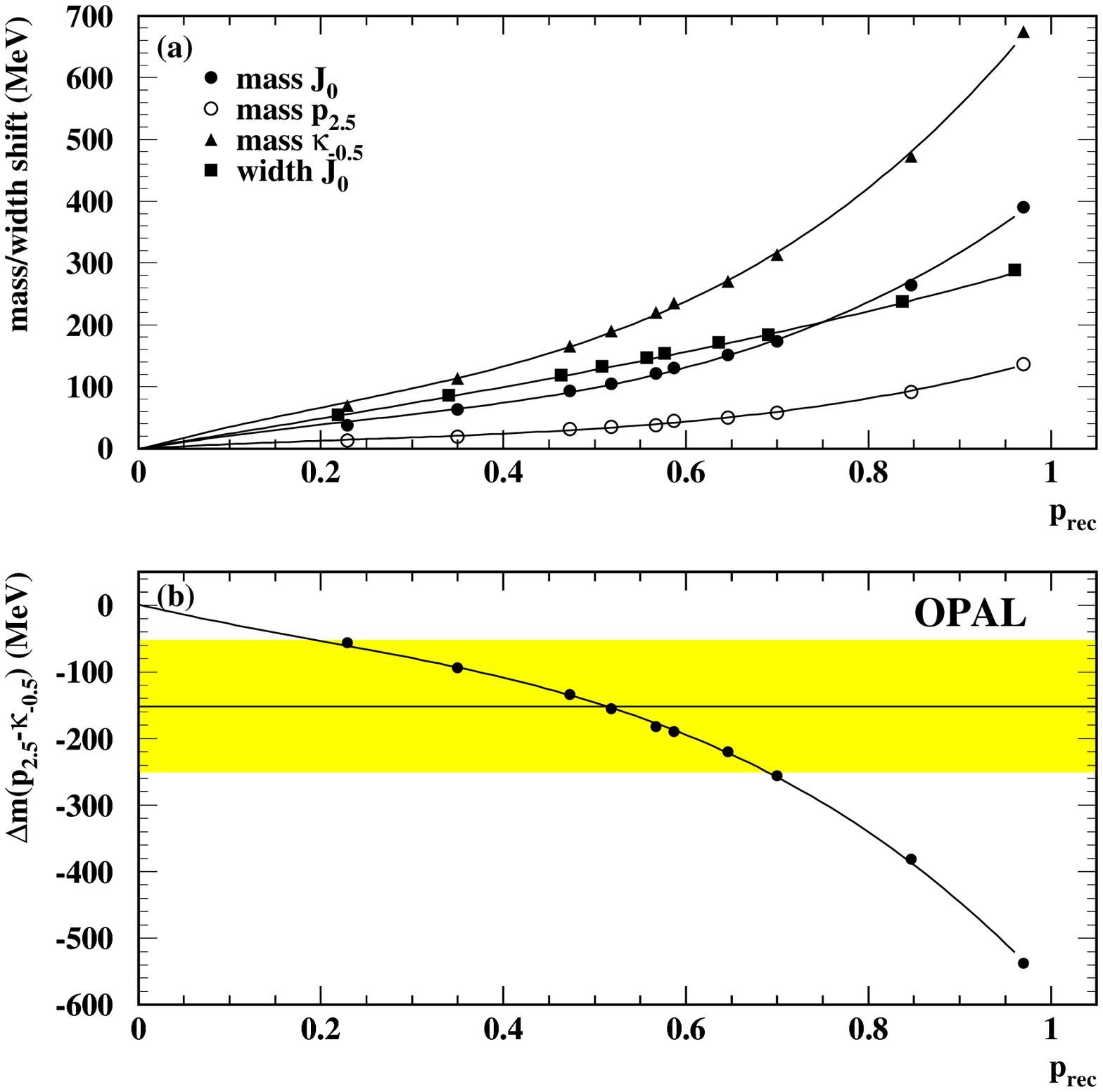}{f:colrec}{(a) W mass and width shifts in the 
\qqqq\ channel convolution fit as a function of reconnection probability 
\precon\ in the SK~I model, for various different jet direction
reconstruction methods (the points for the W width are slightly offset for 
clarity); (b) predicted mass difference $\dmyx{\syaiii}$ as 
a function of reconnection probability \precon. 
The shifts are calculated at the displayed points
using linear interpolation to
$\rs=196$\,GeV, the mean centre-of-mass energy of the data sample, and the
curves are drawn purely to guide the eye.
The value and associated 
error (including both statistical and systematic contributions)
measured from the data are indicated by the horizontal line and shaded band.}

Within the colour reconnection models implemented in {\sc Ariadne},
reconnection occurs if it would reduce the total string length and
is allowed within the constraints of the colour algebra factors 
\cite{crar}. In the first variant (AR1), colour reconnection occurs
only amongst the decay products of one string (from a single W boson),
and not between the two W bosons of a $\ww\rightarrow\qqqq$ event.
In the AR2 model, colour reconnection is also allowed between
strings, and hence between the two W bosons, but only for gluon energies
below 2\,GeV (the natural W width). The mass and width shifts due to
colour reconnection are calculated as the difference between AR2 and AR1, to 
isolate the effect of colour reconnection between W bosons.\footnote{
The 2\,GeV limit on gluon energies for colour reconnection between W
bosons is implemented in AR2 by running the dipole cascade twice,
once down to 2\,GeV with inter-W colour reconnection disabled, then
again with it enabled. This results in some artificial additional 
high-energy showering, which was emulated in AR1 for the purpose
of this comparison by also running the cascade twice with an
interruption at 2\,GeV \cite{lonpriv}.} Finally, in the AR3 model, colour
reconnection is allowed between strings for all gluon energies, 
producing a strong effect and rather large mass and width shifts.
However, this model is disfavoured both theoretically and by 
studies of three-jet events in \zb\ data \cite{ar3dis}. Additionally,
studies of rapidity gaps in three-jet $\zb\rightarrow\qqbar\rm g$
events \cite{rapgap} disfavour the {\sc Ariadne}
implementation of colour reconnection even within one string.

The {\sc Herwig} Monte Carlo program also includes a colour reconnection
model implemented in the framework of cluster hadronisation. 
In this model,
a rearrangement of the association of partons to clusters occurs
with a fixed probability of $1/9$ if this rearrangement would
lead to a smaller space-time extent of the clusters. The resulting 
mass and width shifts, shown in Table~\ref{t:colrec}, are smaller 
than those for the SK~I and {\sc Ariadne}-based models. 

All colour reconnection models studied show mass shifts which are reduced
by factors of two to three by the \syaiii\ jet direction reconstruction method,
and  increased by factors of up to two for the \syavi\ method. Hence,
as discussed in Section~\ref{s:reckin}, 
the \syaiii\ method is used for the main results of the
\qqqq\ channel analysis in this paper. The largest shifts are seen
in the SK~I model (depending on the value of $k_I$), and the
AR2 model (AR3 being disfavoured both theoretically and experimentally).
The final colour reconnection errors for the W mass and width are
discussed in Section~\ref{s:fsilim} below, where constraints from
$\dmyx{\syaiii}$ and studies of particle flow in $\ww\rightarrow\qqqq$
events are also taken into account.

Bose-Einstein correlations (BEC) between like-sign charged pion pairs are
well established in both \zb\ and $\rm W\rightarrow\qqbar$ decays at 
LEP \cite{beclep,opalwwbec,othwwbec}. Although they are not implemented 
in the standard
hadronisation models used in this paper, studies using dedicated
samples generated using the LUBOEI \cite{luboei} BEC model show that 
BEC between decay products originating from the same W boson 
(intra-W BEC) do not
lead to significant W mass or width biases. However, just as in
the case of colour reconnection, BEC between
like-sign particles from different W bosons (inter-W BEC)
may result in significant
mass and width biases in the $\ww\rightarrow\qqqq$ analysis. These 
have been assessed using the LUBOEI BE$_{32}$ model \cite{luboei}
as the difference
between Monte Carlo samples generated with BEC affecting all possible
like-sign particle pairs, and samples with BEC only between pairs from
the same W boson. The parameters governing the properties of the
generated correlations were tuned to describe BEC observed 
in $\zb\rightarrow\qqbar$ decays as described in \cite{opalwwbec},
and were set to $\lambda=2.15$ and $R=0.26$\,GeV. The resulting mass and
width shifts are listed in Table~\ref{t:colrec}. As in the case of
colour reconnection, the mass shifts introduced by inter-W BEC are 
significantly reduced by the \syaiii\ alternative jet direction
reconstruction method, and increased
by the \syavi\ method.

The possible existence of inter-W BEC has been experimentally investigated in
\cite{opalwwbec,othwwbec}, and limits placed on its strength with respect
to that predicted by the LUBOEI model. By fitting the BEC strength
parameter $\Lambda$, the amount of inter-W BEC is measured in \cite{opalwwbec}
to be a fraction  $0.33\pm 0.44$ of that predicted by LUBOEI, corresponding 
to a one standard deviation upper bound of 0.77. Assuming a linear 
relation between the BEC strength and the corresponding W mass and width 
shifts, the systematic errors on the W mass and width shown in 
Tables~\ref{t:mwsyst} and~\ref{t:gwsyst} are set to 77\,\% of those predicted
by the LUBOEI model shown in Table~\ref{t:colrec}.

\subsection{Photon radiation}\label{s:oalpha}

The dominant process contributing to four-fermion final states with 
an additional photon ($\epem\rightarrow\ffbar\ffbar\gamma$) is 
initial-state radiation (ISR) of photons from the incoming electrons and
positrons, where the $O(\alpha^3)$ treatment of 
{\sc KoralW} is of more than adequate precision. 
However, {\sc KoralW} does not 
include all $O(\alpha)$ photon radiation effects, {\em e.g.} radiation
from the W bosons themselves (WSR), and interference between
ISR, WSR and final-state radiation (FSR) from the outgoing charged leptons
and quarks. A more complete treatment
is provided by the so-called {\sc KandY} \cite{kandy}
generator scheme, consisting
of {\sc KoralW} version 1.51 \cite{kandy}
and YFSWW3 \cite{yfsww} running concurrently.
This introduces two major improvements over the {\sc KoralW} 1.41 samples
used to calibrate the mass and width fits, namely the inclusion
of $O(\alpha)$ non-leading electroweak corrections and the screened
Coulomb correction as opposed to the non-screened correction used
previously.

The differences in mass and width biases predicted by {\sc KandY}
and {\sc KoralW} are shown for the convolution fit in 
Table~\ref{t:oasyst}. The results are similar for the other fit
methods, and in the \qqqq\ channel do not depend significantly on
the choice of jet direction reconstruction method.
As {\sc KandY} gives a more complete treatment than {\sc KoralW}, these
shifts are applied as corrections
to the final W mass and width results in this paper.
However, the effects of the  
non-leading electroweak corrections and screened Coulomb corrections
on the W mass partially cancel, and so are considered separately 
when assessing the total systematic error due to photon radiation.

\begin{table}
\centering
\begin{tabular}{l|rr|rr}
 & \multicolumn{2}{c|}{Mass (MeV)} & \multicolumn{2}{c}{Width (MeV)} \\
 & \qqlv & \qqqq & \qqlv & \qqqq \\ \hline
Shifts: & & & & \\
\ {\sc KandY}-{\sc KoralW} & $-2$ & 1 & $-22$ & $-22$ \\
\ No non-leading EW correction & 17 & 13 & 14 & 21 \\
\ No screened Coulomb correction & $-14$ & $-13$ & 15 & 1 \\
\ ISR $O(\alpha)-O(\alpha^3)$ & 1 & 1 & 2 & 1 \\
\hline
Uncertainties: & & & & \\
\ Initial-state radiation & 1 & 1 & 2 & 1 \\ 
\ Non-leading EW corrections & 8 & 6 & 7 & 10 \\
\ Screened Coulomb correction & 7 & 6 & 8 & 1 \\
\ Final state radiation & 1 & 1 & 2 & 2 \\
\hline
Total uncertainty & 11 & 9 & 11 & 10 \\
\end{tabular}
\caption{\label{t:oasyst} Mass and width shifts measured using the 
convolution fit in the \qqlv\ and \qqqq\ channels for various
changes to the treatment of photon radiation (see text). The uncertainties 
due to finite Monte Carlo statistics are 1--2\,MeV. The 
corresponding systematic uncertainties on the W mass and width are also given.}
\end{table}

The main effect of non-leading electroweak corrections is to modify
the ISR spectrum due to ISR-WSR interference \cite{opalwwg}, leading
to bias in the W mass and width analyses, since ISR photons are
not explicitly reconstructed in the kinematic fits. Studies in 
\cite{opalwwg} show that the amount of ISR-WSR interference can be
inferred from the rate of $\ww\gamma$ production with photons at large
angles to the beam direction, and that the data are described best
by a parameter $\kappa=0.38\pm 0.47$, where $\kappa=0$ corresponds
to the treatment in {\sc KandY} and $\kappa=1$ to that in {\sc KoralW}.
The data are therefore consistent with the prediction of {\sc KandY}, and
the uncertainty of 0.47 is used to determine the systematic uncertainties due 
to non-leading electroweak corrections. These are taken to be a fraction $0.47$
of the mass and width shifts induced by reweighting {\sc KandY}
events to remove the corrections (see Table~\ref{t:oasyst}).

The mass and width shifts induced by degrading the screened Coulomb
correction of {\sc KandY} to the unscreened correction implemented
in {\sc KoralW} are also shown in Table~\ref{t:oasyst}. The systematic
uncertainties are taken to be half of these shifts.
Finally, the systematic uncertainty due to the modelling of ISR
is determined by reweighting events to degrade the $O(\alpha^3)$ 
treatment of both {\sc KandY} and {\sc KoralW} to $O(\alpha)$.
The uncertainties due to FSR modelling have been assessed by 
reweighting events to change the rate of FSR from leptons by
$\pm 15$\,\%, and from quarks by $\pm 50$\,\%, based on studies
of \zb\ and W decay data. The mean energies of FSR photons
have also been varied by $\pm 50\,\%$.  
The resulting uncertainties are very small, and also shown in 
Table~\ref{t:oasyst}. The total uncertainties due to photon radiation
are determined from the quadrature sum of all the above sources,
and amount to around 10\,MeV for both the W mass and width.
  
\subsection{Background}\label{s:bgsyst}

In \qqlv\ events, non-WW background is very small, except in the
\qqtv\ channel (see Table~\ref{t:evtsel}). The uncertainty due to
modelling of background from \zg\ events is assessed by using
Monte Carlo samples generated with {\sc Pythia} (which has a simpler
treatment of ISR) instead of KK2f, and by hadronising the KK2f
 samples with various
different hadronisation models, as for signal WW events.
The absolute rates of \zg\ and \zz\ background are varied by
$\pm 20$\,\%, following the uncertainties derived in \cite{wwxsec189}.
The assigned uncertainty for \zz\ background is larger than that for
the on-shell Z-pair production cross-section \cite{opalzz}, but includes
contributions from other $\epem\ffbar$ production diagrams.
The errors assigned for all fit methods 
are given in Tables~\ref{t:mwsyst} and~\ref{t:gwsyst}. The total
background error results from 
approximately equal contributions from hadronisation, \zg\ and ZZ background
rate uncertainties.

Background in the \qqqq\ channel is much larger, and is dominated
by $\epem\rightarrow\zg$ events giving a four-jet final state.
Changing the hadronisation model used for the default KK2f \zg\ 
samples from {\sc Jetset} to  
{\sc Herwig} gives shifts of up to 20\,MeV for the
W mass and 32\,MeV for the W width in the convolution fit, 
with {\sc Pythia} and
{\sc Ariadne} lying in between {\sc Jetset} and {\sc Herwig}.
The Monte Carlo modelling of this
background is further investigated by using four-jet \zb\ decays
taken at $\rs\approx 91$\,GeV, and scaling the energies of all tracks 
and clusters by 200\,GeV$/\mz$ before applying the standard 
$\ww\rightarrow\qqqq$ event selection. These scaled events are
then used in place of the standard \zg\ background samples in the
W mass and width analysis. The mass and width shifts seen when using 
scaled \zb\ events hadronised with {\sc Ariadne} and {\sc Herwig}
instead of {\sc Jetset} reproduce well the shifts seen in \zg\ events in
all analysis methods, although there are some differences
between the scaled \zb\ and \zg\ mass distributions which are sensitive
to details of the selection and scaling procedure. However, in all cases
the scaled four-jet \zb\ data events are found to lie 
between the predictions of {\sc Jetset} and {\sc Herwig}, so
the differences between {\sc Jetset} and {\sc Herwig} \zg\ 
events at high energy are therefore used
to set the systematic errors due to hadronisation in \zg\ events.
Due to the small contribution of $\epem\rightarrow\zz\rightarrow\qqqq$ 
events to the selected sample 
(3--6\,\%) and the similar hadronic properties of $\ww\rightarrow\qqqq$ and 
$\zz\rightarrow\qqqq$ events, the extra hadronisation uncertainty due to 
\zz\ production is neglected.

The absolute rate of \zg\ background events in the \qqqq\ channel
is varied by $\pm 5$\,\%,
based on studies of the modelling of four-jet \zg\ background
described in \cite{wwxsec189} and the modelling of the likelihood
used in the $\ww\rightarrow\qqqq$ selection. The effect of changing
the ISR modelling and ISR-FSR interference in KK2f is checked using
the procedures described in \cite{opalff} and found to be small.
Finally, the absolute rate of $\zz\rightarrow\qqqq$ events is varied
by its uncertainty of $\pm 11$\,\% \cite{opalzz}. The total systematic
errors due to the modelling of non-WW background are given in 
Tables~\ref{t:mwsyst} and~\ref{t:gwsyst}, and are dominated by
uncertainties in the modelling of fragmentation in \zg\ events. 

\subsection{LEP beam energy}

Constraining the total reconstructed energy of the WW decay products
to \rs\ greatly improves the event-by-event W mass resolution, but requires
that the LEP beam energy be precisely known. The latter has been measured
using a combination of resonant depolarisation and magnetic extrapolation
based on NMR probes, complemented by measurements of the total bending field
of the LEP dipoles using a flux loop, measurements of the beam energy
using a dedicated
spectrometer and studies of the accelerator synchrotron tune \cite{lepebeam}.
The beam energy is known to a precision of between 10\,MeV 
and 21\,MeV, the largest
uncertainty applying for the year 2000 data where special techniques were
used to increase the beam energy to the highest possible value \cite{lepebeam}.
The beam energy uncertainties are largely correlated from year to year,
and correspond to uncertainties of around 10\,MeV on the W mass and 3\,MeV
on the width.

Dispersion and other effects introduce a spread of between 160\,MeV and 260\,MeV 
in the event-by-event collision energy \cite{lepebeam}.
This effect, which is not included in 
the Monte Carlo simulations by default, introduces small shifts 
of around 1\,MeV in the W mass and width, the full sizes of which are 
taken as additional systematic errors. A similar shift is produced by 
the average longitudinal boost of the collision centre-of-mass frame at OPAL
of 12--24\,MeV \cite{lepebeam}.
This shift is caused by asymmetries in the distribution of the
LEP radio-frequency accelerating system around the collider ring, 
and its full size is taken as
an additional systematic uncertainty. Both these effects are included in the
LEP beam energy entries of Tables~\ref{t:mwsyst} and~\ref{t:gwsyst}.

\subsection{Other systematic errors}\label{s:other}

Small discrepancies in the modelling of the data by the Monte Carlo 
lead to additional systematic uncertainties in the convolution fit result.
In both channels, the W mass and width bias depend weakly on the fitted
event-by-event error, which on average is up to 1\,\% larger in data than
Monte Carlo. Similarly, there are small discrepancies in the number of
accepted jet assignment combinations in the \qqqq\ channel 
(see Figure~\ref{f:comb}(c)). Both of these effects are assessed from the 
change in fit bias induced by reweighting Monte Carlo distributions to those 
of the data. There are no equivalent uncertainties in the reweighting and
Breit-Wigner fits, where no significant discrepancies are seen in any 
important distributions.

The bias corrections for the convolution and Breit-Wigner fits, and the
template distributions used in the reweighting fit, have small uncertainties
due to finite Monte Carlo statistics. Such uncertainties also play a role
in {\em e.g.} the hadronisation and detector systematics, but are accounted
for in the corresponding systematic errors where appropriate.

\subsection{Consistency checks}

The complete W mass analysis is performed for the convolution, 
reweighting and Breit-Wigner fits, and the width analysis for the convolution
and reweighting fits, with an additional 5C convolution fit in the \qqlv\ 
channel. The results are given in Tables~\ref{t:mresstat} and~\ref{t:wresstat}.
The differences between the various fit results are summarised in 
Table~\ref{t:mcomp}, together with the expected RMS differences evaluated
using a large number of common Monte Carlo simulation subsamples analysed
by each method in a consistent manner. The observed differences between
fit methods are compatible with expectation. The consistency between the
different fit methods, data-taking years and analysis channels can also
be seen in Figures~\ref{f:mresyear} and~\ref{f:wresyear}. Given these
results, no additional systematic uncertainty associated with
any individual fit method is assigned.
Complete systematic error analyses are also performed for all 
fitting methods except the CV5 \qqlv\ width fit
(see Tables~\ref{t:mwsyst} and~\ref{t:gwsyst}), and largely
comparable uncertainties obtained, again giving confidence in the results.

\begin{table}
\centering

\begin{tabular}{l|rrr|rrr}
Channel & \multicolumn{3}{c|}{Fitted \mw} & \multicolumn{3}{c}{Fitted \gw} \\
 & CV$-$RW & RW$-$BW & BW$-$CV & CV$-$RW & RW$-$CV5 & CV5$-$CV \\
\hline
\qqev & $19\pm 48$ & $-8\pm 55$ & $-11\pm 59$ & 
 $-313\pm 170$ & $34\pm 160$ & $279\pm 160$ \\
\qqmv & $-56\pm 52$ & $-35\pm 56$ & $91\pm 68$ & 
 $35\pm 180$ & $8\pm 160$ & $-43\pm 180$ \\
\qqtv & $65\pm 92$ & $20\pm 89$ & $-85\pm 116$ & 
 $-326\pm 260$ & $-115\pm 270$ & $441\pm 290$ \\
\qqlv & $0\pm 33$ & $-6\pm 35$ & $6\pm 42$ & 
 $-162\pm 100$ & $-15 \pm 90$ & $177\pm 90$ \\
\hline 
\qqqq (\syaiii) & $45\pm 45$ & $22\pm 56$ & $-67\pm 54$ & --- & --- & --- \\
\qqqq ($J_0$) & $11\pm 39$ & $-41\pm 43$ & $30 \pm 40$ & $-51\pm 97$ & --- & --- \\
\end{tabular}
\caption{\label{t:mcomp}Differences in fitted W mass and width values
between pairs of fit methods in each analysis channel. The uncertainties
indicate the expected RMS spread in the fitted differences, evaluated
using Monte Carlo simulation subsamples.}
\end{table}

The statistical 
correlations between the fit methods have also been assessed using similar
Monte Carlo subsamples, and found to be between 0.65 and 0.88. Combining the
results from all three methods would only reduce the statistical error on \mw\ 
by around 2\,\%. Given this, and the large correlation between the systematic
uncertainties of the different methods, they are not combined and the 
final results are taken from the convolution fit alone.

\section{Results}\label{s:results}

The final results of the W mass and width fits are presented in this section,
taking into account both statistical and systematic errors. The results for
the measurement of \dmyx{\syaiii}\ are given in Section~\ref{s:fsilim}
and used in conjunction with previous measurements to derive a limit on the
SK~I model parameter $k_I$. The results for the W mass and width incorporating
this limit are then given in Section~\ref{s:nonlres}.

\subsection{Colour reconnection limit}\label{s:fsilim}

The differences in W mass \dmjx\ extracted using an alternative jet direction
reconstruction method  $X$ and
method \syavi\ are sensitive to possible final-state interactions in
the $\ww\rightarrow\qqqq$ channel and can be used to set a limit on colour
reconnection, as discussed in Section~\ref{s:cvres}. Of all the direction
reconstruction methods
considered, Monte Carlo studies show that the mass difference \dmyx{\syaiii}\
is the most sensitive to the SK~I model, for both moderate and large amounts
of colour reconnection, and is therefore used to set a limit on $k_I$.

The systematic uncertainties on \dmyx{\syaiii}\ are evaluated using the same
techniques as discussed in Section~\ref{s:syst} and are given in 
Table~\ref{t:mwsyst}. Detector effects are largely correlated between 
jet algorithms and the resulting residual uncertainties are very small.
Hadronisation effects are calculated by taking the effect on \dmyx{\syaiii}\ 
when simultaneously changing the hadronisation models used to set the bias 
corrections in both mass measurements, and are dominated by the difference
between JT and AR$'$ (see Table~\ref{t:hadrm}). 
Background and modelling uncertainties are assessed
similarly. The largest uncertainty comes from Bose-Einstein 
correlations, which have a similar effect to colour reconnection on the
value of 
\dmyx{\syaiii}, albeit with the opposite sign. 
Taking all uncertainties into account, the result is:
\[
\dmyx{\syaiii} =  \dmres \pm \dmstat \pm \dmsyst \pm \dmbec\rm\, MeV,
\]
where the first error is statistical, the second systematic excluding
Bose-Einstein correlations and the third from Bose-Einstein correlations.
Within
the SK~I model, this corresponds to a value of $k_I=1.7^{+2.0}_{-1.2}$,
or a colour reconnection probability of $\precon=0.51^{+0.17}_{-0.27}$ at
$\sqrt{s}$=196\,GeV (see Figure~\ref{f:colrec}).

This result is combined with the OPAL limit determined from studies
of particle flow in inter-jet regions of $\ww\rightarrow\qqqq$ events
\cite{opflow}.\footnote{The analysis of W decay charged multiplicity 
differences between $\ww\rightarrow\qqlv$ and $\ww\rightarrow\qqqq$
events presented in \cite{opflow} is not sensitive enough to colour
reconnection to make any significant difference to the combination
presented here.} Statistical correlations between the two methods were 
evaluated using Monte Carlo subsample techniques, and found to be less
than 10\,\%. The largest systematic errors from hadronisation and Bose-Einstein
correlation were taken to be fully correlated. The final result
is $\precon=0.43^{+0.15}_{-0.20}$, with the one standard deviation 
upper bound of
$\precon<0.58$ corresponding to $k_I=2.3$. This value of $k_I$ has been
used to calculate the systematic errors due to colour reconnection
for the W mass and width shown in Tables~\ref{t:mwsyst} and~\ref{t:gwsyst},
based on the SK~I model. The systematic errors which would result for
other values of $k_I$ can be seen in Figure~\ref{f:colrec} and are listed
in Table~\ref{t:crlim}.
The corresponding limit at 95\,\% confidence level
is $\precon<0.70$ or $k_I<4.0$. The particle flow and
$\dmyx{\syaiii}$ results are not sensitive enough to test the predictions
of the AR2 and {\sc Herwig} colour reconnection models, but these give
mass and width shifts smaller than those predicted by the SK~I model
with $k_I=2.3$, as can be seen from Table~\ref{t:colrec}. 

The final W mass and width fit results are not adjusted to compensate for
the possible effects of colour reconnection.
A constant colour reconnection uncertainty
is assumed, independent of \rs\ and evaluated at the mean
centre-of-mass energy of the data sample. This avoids changing the relative
weights of the different energy points as a function of the energy 
dependence of the colour reconnection uncertainty of any particular model.

\begin{table}
\centering

\begin{tabular}{cc|cc}
$k_I$ & \precon & \multicolumn{2}{c}{Uncertainty} \\
& & Mass & Width \\
\hline
0.9 & 0.35 & 19 & 86 \\
1.5 & 0.47 & 31 & 119 \\
1.8 & 0.52 & 35 & 133 \\
2.3 & 0.58 & 41 & 151 \\
3.1 & 0.65 & 50 & 172 \\
\end{tabular}
\caption{\label{t:crlim}Systematic uncertainties (in MeV) on the \qqqq\ 
channel measurements of the W mass and width
in the SK~I model for various values of the model parameter $k_I$ and 
associated  colour reconnection probability \precon. The uncertainties
corresponding to $k_I=2.3$ are used for the final result.}
\end{table}

\subsection{Results for the W mass and width}\label{s:nonlres}

The results for the W mass and width, including both statistical and 
systematic errors, are given for all analysis methods in 
Table~\ref{t:ressum}. The central values differ slightly from those in 
Tables~\ref{t:mresstat} and~\ref{t:wresstat} as systematic errors have been 
taken into account in the combination of results from different years.
Combining both \qqlv\ and \qqqq\ channels using
a $\chi^2$ minimisation technique, and assuming all systematic errors to be
correlated between the two channels, the final results 
(taken from the convolution fit and using the running width scheme for the
Breit-Wigner distribution as implemented in {\sc KoralW}) are:
\begin{eqnarray*}
\mw (\qqlv+\qqqq) & = & \mwnonl \pm \mwnonlstat \pm \mwnonlsyst \pm 
\mwnonllep\,\rm GeV\,, \\
\gw (\qqlv+\qqqq) & = &  \gwnonl \pm \gwnonlstat \pm \gwnonlsyst \pm 
\gwnonllep \rm\,GeV\,,
\end{eqnarray*}
where in each case the first error is statistical, the second systematic and 
the third due to the uncertainty in the LEP beam energy. The 
estimated correlation between the two results is 0.04. The full breakdown
of systematic errors is given in Tables~\ref{t:mwsyst} and~\ref{t:gwsyst}.
The \qqqq\ channel has a weight of 0.34 in the mass measurement and 0.35 in the
width measurement; the \qqqq\ weights for the other fitting techniques
are similar. In the absence of systematic uncertainties due to final-state 
interactions, and using the $J_0$ jet direction reconstruction method,
the statistical error of the combined W mass measurement would be 
0.038\,GeV, only
10\,\% smaller than the \mwnonlstat\,GeV of the present result. 
This demonstrates that the modified
jet direction reconstruction technique significantly reduces uncertainties
due to final-state interactions, and allows most of the statistical power
of the \qqqq\ channel to be exploited.

\begin{table}
\centering

\begin{tabular}{l|ccc}
& \multicolumn{3}{c}{W mass (GeV)} \\
& Convolution & Reweighting & Breit-Wigner \\ \hline
\qqlv & $\mwqqlv\pm\mwqqlvstat\pm\mwqqlvsytt$ & 
$80.451\pm 0.058\pm 0.022$ & 
$80.457\pm 0.063\pm 0.029$ \\
\qqqq (\syaiii) & $\mwqqqq\pm\mwqqqqstat\pm\mwqqqqsytt$ & 
$80.308\pm 0.064\pm 0.056$ &
$80.278\pm 0.072\pm 0.057$ \\
\qqqq ($J_0$) &  $\it\mwqqjz\pm\mwqqjzstat\pm\mwqqjzsytt$ & 
$\it 80.383\pm 0.056\pm 0.136$ &
$\it 80.416\pm 0.058\pm 0.137$ \\
\hline
Combined & $\mwnonl\pm\mwnonlstat\pm\mwnonlsytt$ & 
$80.405\pm 0.044\pm 0.028$ &
$80.390\pm 0.048\pm 0.032$ \\
 & \\
& \multicolumn{3}{c}{W width (GeV)} \\
& Convolution & Reweighting \\ \hline
\qqlv & $\gwqqlv\pm\gwqqlvstat\pm\gwqqlvsytt$ & 
$2.088\pm 0.131\pm 0.085$ \\
\qqqq ($J_0$) & $\gwqqqq\pm\gwqqqqstat\pm\gwqqqqsytt$ &
$2.176\pm 0.130\pm 0.180$  \\
\hline
Combined & $\gwnonl\pm\gwnonlstat\pm\gwnonlsytt$ & 
$2.113\pm 0.101\pm 0.097$  \\
\end{tabular}
\caption{\label{t:ressum}Summary of W mass and width results for all fit
methods. In each case, the first error is statistical and the second
systematic, including the error on the LEP beam energy.
The \qqqq\ mass results with the $J_0$ jet direction reconstruction algorithm
are shown for comparison purposes, and are not included in the combination.
The results quoted for the Breit-Wigner fit include the 172\,GeV results
from \cite{wmass172}, as these data have not been reanalysed using this fit
method.} 
\end{table}

The differences between the fitted values of \mw\ and \gw\ in the \qqqq\
and \qqlv\ channels are:
\begin{eqnarray*}
\Delta\mw (\qqqq-\qqlv) & = & \dmwc \pm \dmwcstat \pm \dmwcsyst\,\rm GeV\,, \\
\Delta\gw (\qqqq-\qqlv) & = & \dgwc \pm \dgwcstat \pm \dgwcsyst\,\rm GeV\,, 
\end{eqnarray*}
where in each case
the first error is statistical and the second systematic, excluding
uncertainties due to possible final-state interactions in the \qqqq\ channel.
For these results, the hadronisation and background uncertainties are 
conservatively taken to be uncorrelated
between the \qqlv\ and \qqqq\ channels, and all other uncertainties are taken
to be fully correlated.
Significant non-zero values of $\Delta\mw$ or $\Delta\gw$ 
could indicate that final-state interactions are biasing the values determined
from the \qqqq\ channel; however these values are consistent both with zero
and with the shifts expected from colour reconnection in the SK~I model
with $k_I=2.3$.

The result for the W mass is combined with previous OPAL measurements
using $\ww\rightarrow\lvlv$ events at \rs\ values between 183\,GeV and 
209\,GeV \cite{mwlvlv} and from
the dependence of the WW production cross section on \mw\ at 
$\rs\approx 161$\,GeV \cite{mwthres}. The final OPAL result for the 
W mass is then
\[
\mw=\mwallo \pm \mwallostat \pm \mwallosyst \pm \mwallolep\,\rm GeV\,,
\]
where again the first error is statistical, the second systematic
and the third due to uncertainties in the LEP beam energy.

\section{Conclusions}\label{s:conc}

The mass and width of the W boson are measured using 
$\epem\rightarrow\ww$ events from the complete 
data sample collected by OPAL at centre-of-mass energies between 170\,GeV and
209\,GeV, using event-by-event reconstruction of the W mass in 
the \qqlv\ and \qqqq\ final states. The result for \mw\ is combined
with earlier OPAL results using \lvlv\ events \cite{mwlvlv} and the dependence
of the WW production cross-section on \mw\ at threshold \cite{mwthres}.
The final results are:
\begin{eqnarray*}
\mw & = & \mwallo \pm \mwallostat \pm \mwallosyst \pm \mwallolep \rm\,GeV\,,\\
\gw & = &  \gwnonl \pm \gwnonlstat \pm \gwnonlsyst \pm 
\gwnonllep \rm\,GeV\,,
\end{eqnarray*}
where the first error is statistical, the second systematic and the
third due to the uncertainty on the LEP beam energy. These results
are consistent with, and supersede, our previous results 
\cite{wmass172,wmass183,wmass189}, and are also consistent with
other values measured at LEP \cite{mgwlep} and the Tevatron \cite{mgwtev}.
The results are also consistent with the values inferred indirectly from
precision electroweak data \cite{smew}, providing a powerful 
consistency test of the Standard Model of electroweak interactions.

Limits are placed on the strength of possible final-state interactions
in $\epem\rightarrow\ww\rightarrow\qqqq$ events, by studying the evolution
of the fitted W mass measured with various jet algorithms having differing 
sensitivities to such interactions. In combination with OPAL results based
on particle flow in the regions between jets in 
$\epem\rightarrow\ww\rightarrow\qqqq$
events, a 95\,\% confidence level upper limit on the SK~I colour reconnection
model parameter $k_I$ is set at $k_I<4.0$, corresponding to a colour
reconnection probability of 0.70.

\section*{Acknowledgements}

We particularly wish to thank the SL Division for the efficient operation
of the LEP accelerator at all energies
 and for their close cooperation with
our experimental group.  In addition to the support staff at our own
institutions we are pleased to acknowledge the  \\
Department of Energy, USA, \\
National Science Foundation, USA, \\
Particle Physics and Astronomy Research Council, UK, \\
Natural Sciences and Engineering Research Council, Canada, \\
Israel Science Foundation, administered by the Israel
Academy of Science and Humanities, \\
Benoziyo Center for High Energy Physics,\\
Japanese Ministry of Education, Culture, Sports, Science and
Technology (MEXT) and a grant under the MEXT International
Science Research Program,\\
Japanese Society for the Promotion of Science (JSPS),\\
German Israeli Bi-national Science Foundation (GIF), \\
Bundesministerium f\"ur Bildung und Forschung, Germany, \\
National Research Council of Canada, \\
Hungarian Foundation for Scientific Research, OTKA T-038240, 
and T-042864,\\
The NWO/NATO Fund for Scientific Research, the Netherlands.

\end{document}